\documentclass[a4paper,fleqn,usenatbib]{mnras} 

\usepackage{newtxtext,newtxmath}

\usepackage[T1]{fontenc}
\usepackage{ae,aecompl}

\usepackage{graphicx}	
\usepackage{amsmath}	
\usepackage{amssymb}	

\newcommand{\bs}[1]{{\mbox{\boldmath $#1$}}}

\title[Rising flux tube and formation of sunspots]{On rising 
magnetic flux tube and formation of sunspots in a deep domain}

\author[H. Hotta \& H. Iijima]{
H. Hotta,$^{1}$\thanks{E-mail: hotta@chiba-u.jp}
H. Iijima,$^{2}$
\\
$^{1}$Department of Physics, Graduate School of Science, Chiba
university, 1-33 Yayoi-cho, Inage-ku, Chiba, 263-8522, Japan\\
$^{2}$Division for Integrated Studies, Institute for Space-Earth 
Environmental Research, Nagoya university, Furocho, Chikusa-ku, 
Nagoya, Aichi 464-8601, Japan\\
}

\date{Accepted XXX. Received YYY; in original form ZZZ}

\pubyear{}

\begin{document}
\label{firstpage}
\pagerange{\pageref{firstpage}--\pageref{lastpage}}
\maketitle

\begin{abstract}
  We investigate the rising flux tube and the formation of sunspots in an unprecedentedly deep computational domain that covers the whole convection zone with a radiative magnetohydrodynamics simulation. Previous calculations had shallow computational boxes ($<30$ Mm) and convection zones at a depth of 200 Mm. By using our new numerical code R2D2, we succeed in covering the whole convection zone and reproduce the formation of the sunspot from a simple horizontal flux tube because of the turbulent thermal convection. The main findings are (1) The rising speed of the flux tube is larger than the upward convection velocity because of the low density caused by the magnetic pressure and the suppression of the mixing. (2) The rising speed of the flux tube exceeds 250 m/s at a depth of 18 Mm, while we do not see any clear evidence of the divergent flow 3 hr before the emergence at the solar surface. (3) Initially, the root of the flux tube is filled with the downflows and then the upflow fills the center of the flux tube during the formation of the sunspot. (4) The essential mechanisms for the formation of the sunspot are the coherent inflow and the turbulent transport. (5) The low-temperature region is extended to a depth of at least 40 Mm in the matured sunspot, with the high-temperature region in the center of the flux tube. Some of the findings indicate the importance of the deep computational domain for the flux emergence simulations.
\end{abstract}



\section{Introduction}
Sunspots are one of the most prominent phenomena at the 
solar surface. Sunspots have a strong magnetic field ($>3000$ G), 
which causes suppression of the convective heat transport and
resultant darkening 
\citep[see review by][]{2011LRSP....8....4B}.
The origin of sunspots is thought to be deep in the convection
zone \citep{1955ApJ...121..491P}. The 
magnetic flux tube generated by the dynamo action
is thought to rise to the surface and form sunspots
at the solar surface \citep{1985SoPh..100..397Z}.\par
To understand the rising of the flux tube
and the formation of sunspots using numerical
simulations, realistic physics 
processes
such as
radiation, ionization, and thermal convection
need to be considered. All of these are fundamentally important, and several numerical simulations of the formation
of sunspots have been carried out using these 
processes.
The first numerical simulation for sunspot formation 
including all these
process
was performed by \cite{2010ApJ...720..233C}
using a 7.5-Mm deep computational box.
The magnetic flux torus is inserted from the bottom boundary
kinematically with a certain velocity. They find  that the primary
mechanism of the formation of sunspots is the turbulent
correlation against the mean diverging motion in the central region.\par
\cite{2014ApJ...785...90R} 
performed simulations of emerging magnetic flux with a 15-Mm deep calculation domain.
They find that the continuous
upflow at the bottom boundary prevents the formation of sunspots at the
photosphere; thus, they needed to change the boundary condition from the
forced upflow to free open during the formation process. Later,
\cite{2016SciA....2E0557B} compare their helioseismic observations
of the surface divergent flow with 
\cite{2014ApJ...785...90R}'s calculations with different inserted velocities.
Their observations do not find any clear evidence of the divergent flow at
3 hr
before the emergence time. 
In the simulations, the divergence flow 3 hr before the emergence time is seen when the inserted velocity is large at the bottom boundary at a depth of 18 Mm.
They conclude that the rising velocity should 
not be larger than 150 $\mathrm{m\ s^{-1}}$ at the bottom boundary not to
have clear evidence of the divergent flow against the fluctuation 
caused by supergranulation.
\par
\cite{2012ApJ...753L..13S} performed a sunspot simulation
with a 20-Mm box in the vertical direction.
The initial magnetic condition is the homogeneous flux sheet, and
the turbulent convection spontaneously generates the flux rope and 
a resulting sunspot pair at the solar surface.\par
In these simulations, they need to assume the initial condition of the 
magnetic flux tube or sheet. To exclude this 
voluntariness,
\cite{2017ApJ...846..149C} adopt the bottom boundary condition referred
from dynamo calculation \citep{2014ApJ...789...35F}.
There have been a number of dynamo calculations in which the large-scale 
magnetic field and cycle are reproduced 
\citep{2010ApJ...715L.133G,2012ApJ...755L..22K,2013ApJ...762...73N,2016Sci....351..1427}.
None of these studies include the photosphere, and the typical top boundary is
30 Mm below the photosphere. \cite{2014ApJ...789...35F} find spontaneous flux rising
in their simulation, but the rising scale is about 800 Mm, which is much larger than 
what is expected from the photospheric observation. Thus, \cite{2017ApJ...846..149C}
rescale the results from the dynamo calculation by a factor of 4--8 in space 
to fit their photospheric calculation. The time scale and rising speed
are also changed from the original dynamo calculation.
In addition, the boundary condition is not influenced by what is occurring in the
photospheric calculation. 
They find deep-seated downflow at the bottom boundary with the monolithic structure
of the generated sunspot. They find that the converging flow accompanying the downflow collects the magnetic flux. 
\par
While the understanding of the formation of sunspots has been
significantly
improved in the last decade because of the realistic simulations presented,
the computational domains of these studies are 
relatively shallow (<30 Mm) 
compared with the depth of the convection zone (200 Mm).
We would expect some boundary effects on the resulting
evolution of the flux tube and sunspots.
To minimize the boundary influence to the evolution of the flux tube
and a generated sunspot, we need to extend the calculation box. Recently, our
new numerical code R2D2 (
Radition and RSST for deep dynamics, where RSST is the reduced speed of sound technique
) succeeds in covering the whole convection zone in a calculation \citep{2019SciA....eaau2307}.
In this study, we 
carry out a calculation
of the rising flux tube with an unprecedentedly 
deep calculation box that covers the whole convection zone.  Recently, \cite{2019ApJ...886L..21T}
perform a flux emergence simulation with a 140-Mm deep calculation box using the R2D2
code. In that study, we find spontaneous formation of the delta-type sunspot, which
tends to have solar flares. In this study, we investigate the detailed mechanism of the 
rising of the flux tube and the formation of the sunspot in a deep domain in which 
the influence of the bottom boundary is expected to be small.\par
The rest of the paper is structured as follows. Section \ref{model} describes the
scheme and the setting of the numerical simulation. Section \ref{result} shows the 
calculation results of the rising flux tube, formation process, and 
structure of the matured sunspot. In Section \ref{summary}, we summarize our
results and discuss the differences from previous studies. Future
perspectives of flux emergence simulations are also discussed.

\section{Model}\label{model}
\subsection{Equations}
We solve the three-dimensional magnetohydrodynamic (MHD) equations 
in the Cartesian geometry
$(x,y,z)$ with the radiation transfer, where the $z$-direction is vertical
and the $x$ and $y$ directions are horizontal.
In this study, $z=0$ is the solar surface, which is
$R_\odot=696\ \mathrm{Mm}$ from the center of the sun.
The equations are solved with the R2D2 code.
The magnetohydrodynamic equations are expressed as:
\begin{eqnarray}
 \frac{\partial \rho_1}{\partial t} &=& -\frac{1}{\xi^2}
\nabla\cdot\left(\rho \bs{v}\right), \\
\frac{\partial}{\partial t}\left(\rho \bs{v}\right) &=&
-\nabla\cdot\left(\rho \bs{vv}\right) - \nabla p_1 - \rho_1 g\bs{e_z}
+ \frac{1}{4\pi}\left(\nabla\times\bs{B}\right)\times\bs{B}, \\
\frac{\partial \bs{B}}{\partial t} &=& \nabla\times\left(\bs{v}
\times\bs{B}\right),\\
\rho T \frac{\partial s_1}{\partial t} &=&
\rho T \left(\bs{v}\cdot\nabla\right) s + Q,\label{entropy}\\
p_1 &=& p_1\left(\rho,s\right),
\end{eqnarray}
where $\rho$, $\bs{v}$, $\bs{B}$, $p$, $T$, $s$, $g$, and $Q$ are the 
density, fluid velocity, magnetic field, gas pressure,
temperature, entropy,
gravitational acceleration, and radiative heating, respectively. 
The subscript 1 indicates the perturbation from the stationary
1D stratification indicated with
subscript 0. Thus, the thermodynamic variables are expressed as:

\begin{eqnarray}
  \rho &=& \rho_0 + \rho_1, \\
  p &=& p_0 + p_1, \\
  s &=& s_0 + s_1.
\end{eqnarray}

In this study, we do not assume that the perturbation is smaller 
than the stationary background stratification, i.e., 
$\rho_1 << \rho_0$ is not assumed.
The background stratification is calculated with the hydrostatic 
equation with the help of the Model S
\citep{1996Sci...272.1286C}. See Appendix \ref{background} 
for the details of the calculation procedure.
\par
The equations are solved with the fourth-order spatial derivative
(see Appendix \ref{ihomogeneous_grid} for details)
and the four-step Runge--Kutta method for the time integration 
\citep{2005A&A...429..335V}. To maintain
the stability of the calculation, we adopt the slope-limited diffusion suggested by \cite{2014ApJ...789..132R}.
\par
We use the equation of 
state
considering the partial ionization 
effect with the OPAL repository \citep{1996ApJ...456..902R}.
We switch the linear and table equations of 
state
to address the 
significant change of the perturbation through the convection zone.
We evaluate value

\begin{eqnarray}
 c_\mathrm{eos} = \max 
\left(
 \frac{|s_1|}{s_0},\frac{|\rho_1|}{\rho_0}
\right).
\end{eqnarray}
If $c_\mathrm{eos}$ exceeds $10^{-2}$, we use the table equation of 
state
; otherwise, the linear equation of state,
\begin{eqnarray}
 p_1 = \left(\frac{\partial p}{\partial \rho}\right)_s    \rho_1 
     + \left(\frac{\partial p}{\partial    s}\right)_\rho s_1,
\end{eqnarray}
is used. $(\partial p/\partial \rho)_s$ and 
$(\partial p/\partial s)_\rho$ are prepared with 
the background stratification and only depend on the height ($z$). 
Regarding the table equation of
state, we prepare $64\times64$ grid 
on the density and entropy.
\par
\begin{figure}
  \centering
  \includegraphics[width=0.5\textwidth,pagebox=artbox]{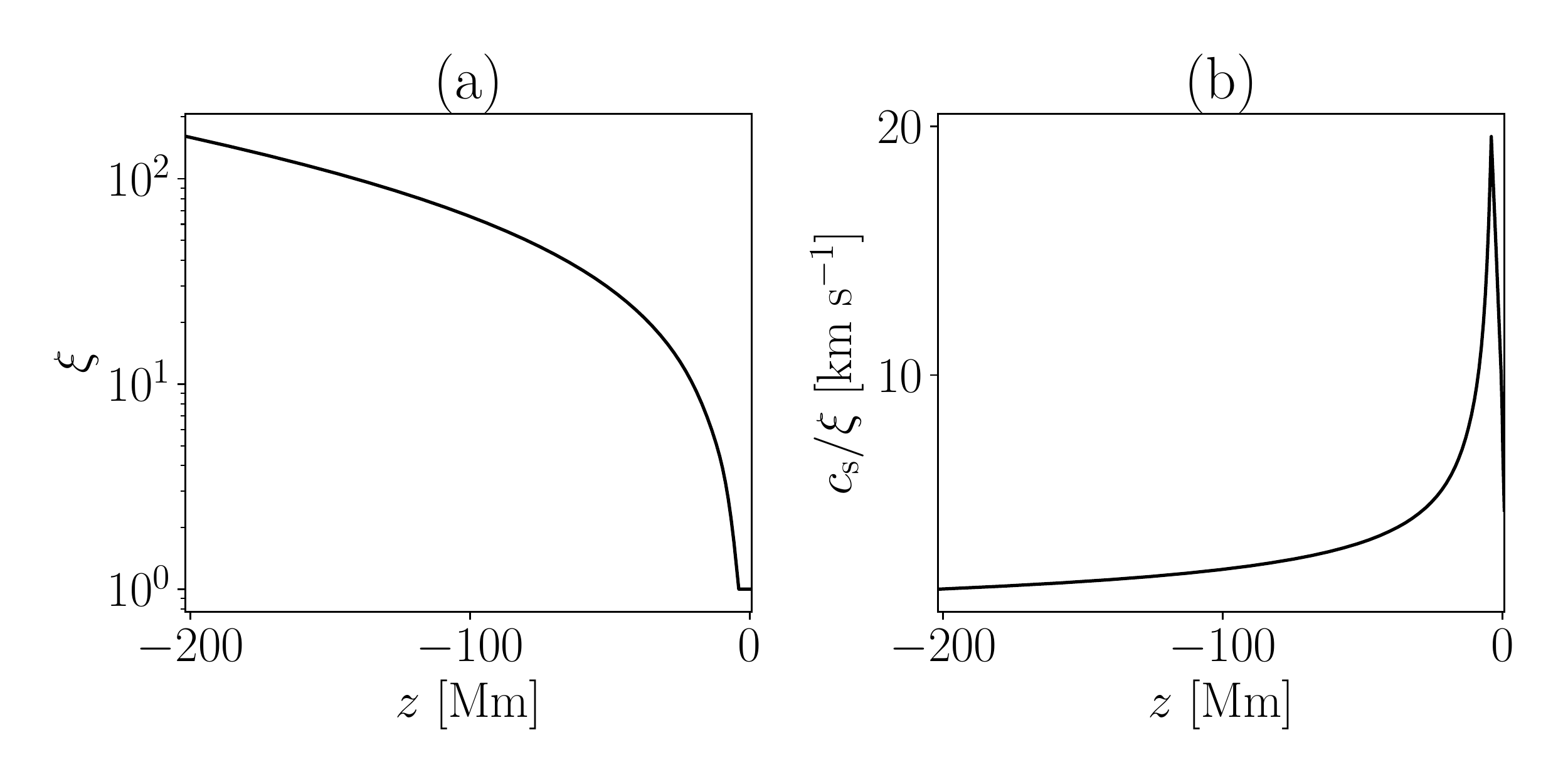}
  \caption{
    (a) The RSST factor $\xi$ and (b) the reduced speed of sound $c_\mathrm{s}/\xi$ are shown.
    }
  \label{xi}
\end{figure}

We adopt the RSST
\citep{2012A&A...539A..30H,2015ApJ...798...51H,2019A&A...622A.157I} for the equation
of continuity to deal with the low Mach number flow in 
the deep convection zone.
By using the RSST, we try to keep the Mach number throughout
the convection zone. To this end, we set the RSST factor,
\begin{eqnarray}
  \xi(z) = \max \left(1,\xi_0
  \left[\frac{\rho_0(z)}{\rho_\mathrm{b}}\right]^{1/3}
  \frac{c_\mathrm{s}(z)}{c_\mathrm{b}}\right),
\end{eqnarray}
Here, we adopt $\xi_0=160$. $\rho_\mathrm{b}=0.2 \mathrm{\ g\ cm^{-3}}$ 
and $c_\mathrm{b}=2.2\times10^7\ \mathrm{cm\ s^{-1}}$ are the density 
and the speed of sound around the base of the convection zone,
respectively.
$c_\mathrm{s}=\sqrt{\left(\partial p/\partial \rho\right)_s}$
is the local adiabatic
speed of sound.
When the energy flux is fixed, the convection velocity scales with 
$\rho_0^{-1/3}$ in the mixing length theory.
This setting approximately maintains the Mach
number. 
Fig. \ref{xi} shows the RSST factor $\xi$ (panel a) and the reduced speed of sound $c_\mathrm{s}/\xi$ (panel b). \cite{2012A&A...539A..30H} show that if the Mach number estimated with the reduced speed of sound is smaller than 0.7, the RSST cause no side effect on the result.
\par
We limit the Alfven velocity to 40 $\mathrm{km\ s^{-1}}$ to deal with low-$\beta$ region above the photosphere \cite{2009ApJ...691..640R}. This does not affect what is occurring at the photosphere.
\par
The radiative heating $Q$ is calculated with
the radiative transfer equation; the details are shown in 
Appendix \ref{radiation}. \par
We calculate the thermal convection with the top boundary at 
$z=-7\ \mathrm{Mm}$ for 90 days.
The thermal convection around the photosphere is very fast, and 
we exclude this layer for accelerating
the calculation. Then, we include the photosphere with the top boundary 
at 700 km above the photosphere and 
continue the calculation for 5 days. This procedure is justified
because the existence of the photosphere
does not change the deep structure
\citep{2019SciA....eaau2307}.

\par
The calculation domain extends 98.304 Mm horizontally for
$x$ and $y$ directions.
The vertical calculation extent is from the base of the convection
zone $z=-0.29R_\odot$ and to $z=700\ \mathrm{km}$.
The number of grid points in each horizontal direction is 1024,
and the grid spacing is 96 km, and this is acceptable to resolve 
the photosphere. In the vertical direction, we use
512 grid points and nonuniform grid spacing, and this is
48 km around the photosphere and 900 km
around the base of the convection zone.
\subsection{Initial condition}\label{flux_tube_setting}
\begin{figure}
  \centering
  \includegraphics[width=0.5\textwidth,pagebox=artbox]
  {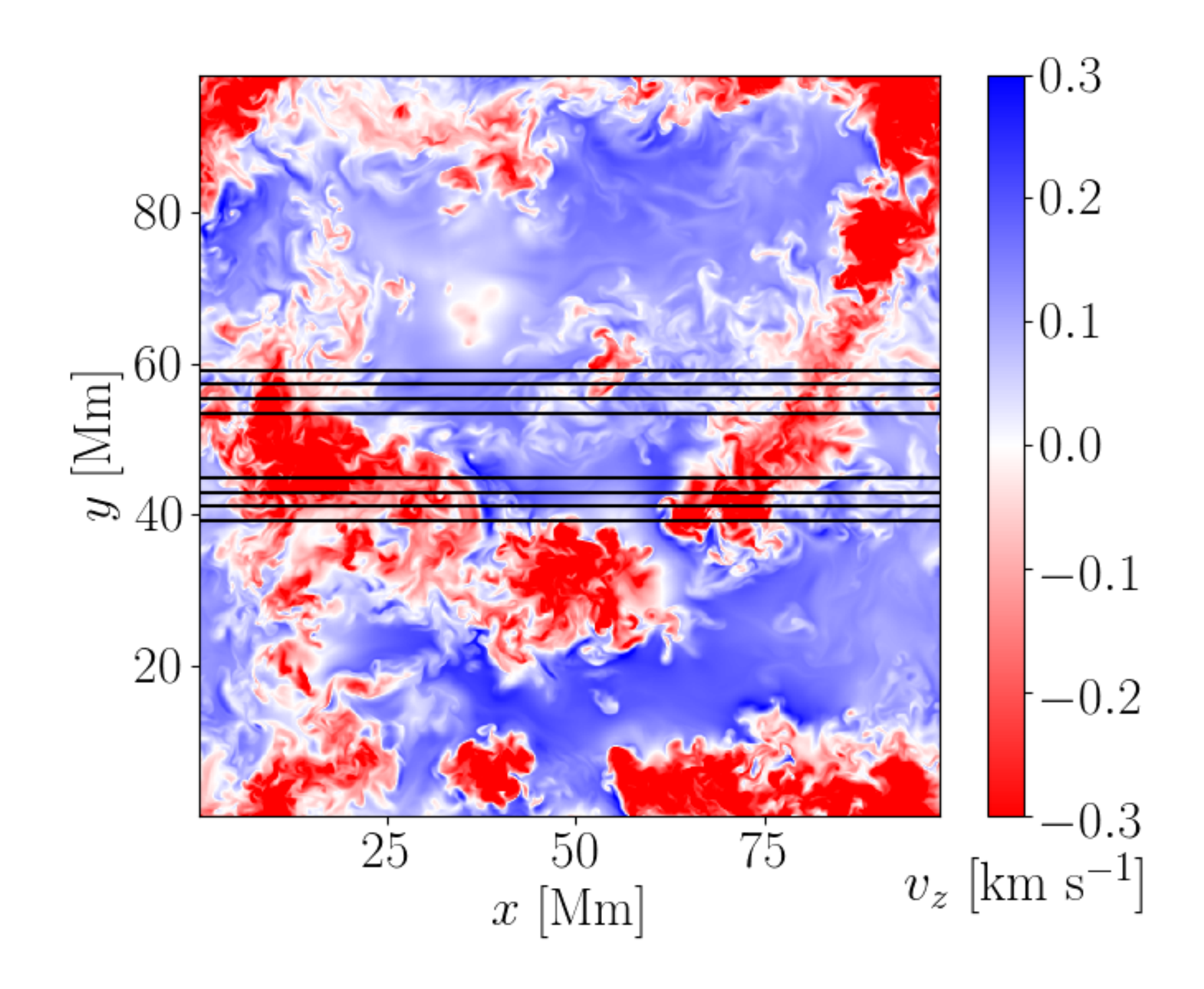}
  \caption{ Initial setup of the magnetic flux tube at $z=-35\ \mathrm{Mm}$. 
 The color contour shows the vertical velocity $v_z$, and
 the contour lines show the 
 axial magnetic field $B_x$. Each contour line shows 0, 2.5, 5, 7.5 kG.
   \label{initial_condition}}
\end{figure}

 We adopt the Lundquist solution \citep{10003639556} for 
 the linear force-free flux tube as follows:
\begin{eqnarray}
  B_x(r) &=& B_\mathrm{tb}J_0(\alpha r), \\
  B_z(y,z) &=&  B_\theta(r)\frac{y-y_\mathrm{tb}}{r}, \\
  B_y(y,z) &=& - B_\theta(r)\frac{z-z_\mathrm{tb}}{r}, \\
  B_\theta(r) &=& B_\mathrm{tb}J_1(\alpha r), \\
  r &=& \sqrt{(y-y_\mathrm{tb})^2 + (z-z_\mathrm{tb})^2 }, \\
 \alpha &=& \frac{a_0}{r_\mathrm{tb},}
\end{eqnarray}
where $a_0 = 2.404825$ is the first root of $J_0$, and
$r_\mathrm{tb}=9.9\ \mathrm{Mm}$ is the radius
of the flux tube. $J_0$ and $J_1$ are the Bessel functions.
Here, we adopt $z_\mathrm{tb}=-0.05R_\odot$, i.e., the initial
magnetic flux tube is located
about 35 Mm below the photosphere and 
$B_\mathrm{tb}=10^4\ \mathrm{G}$ for the magnetic field strength
at the center of the flux tube.
This leads to a total flux of $1\times10^{22}\ \mathrm{Mx}$.
$y_\mathrm{tb}$ is an arbitrary
parameter with which the horizontal
location of the flux tube is determined. With our
choice, the initial flux tube is located in the 
center of the computational domain. 
Fig. \ref{initial_condition} shows the initial flux tube condition
at $z=-0.05R_\odot$. The initial flux tube is caught 
by the two coherent downflows. Note that 
the left downflow is larger and more coherent.
Because the initial magnetic flux tube is force-free, we do not change any
thermodynamic variable, i.e., the density, pressure, or entropy from
the prepared hydrodynamic calculation. Thus, the magnetic flux initially does not
have any buoyancy by the magnetic field, but the inertia of the convective flow
can distort the magnetic flux tube. 

\section{Results}\label{result}
\subsection{Magnetic flux and area}
\begin{figure}
  \centering
  \includegraphics[width=0.5\textwidth]{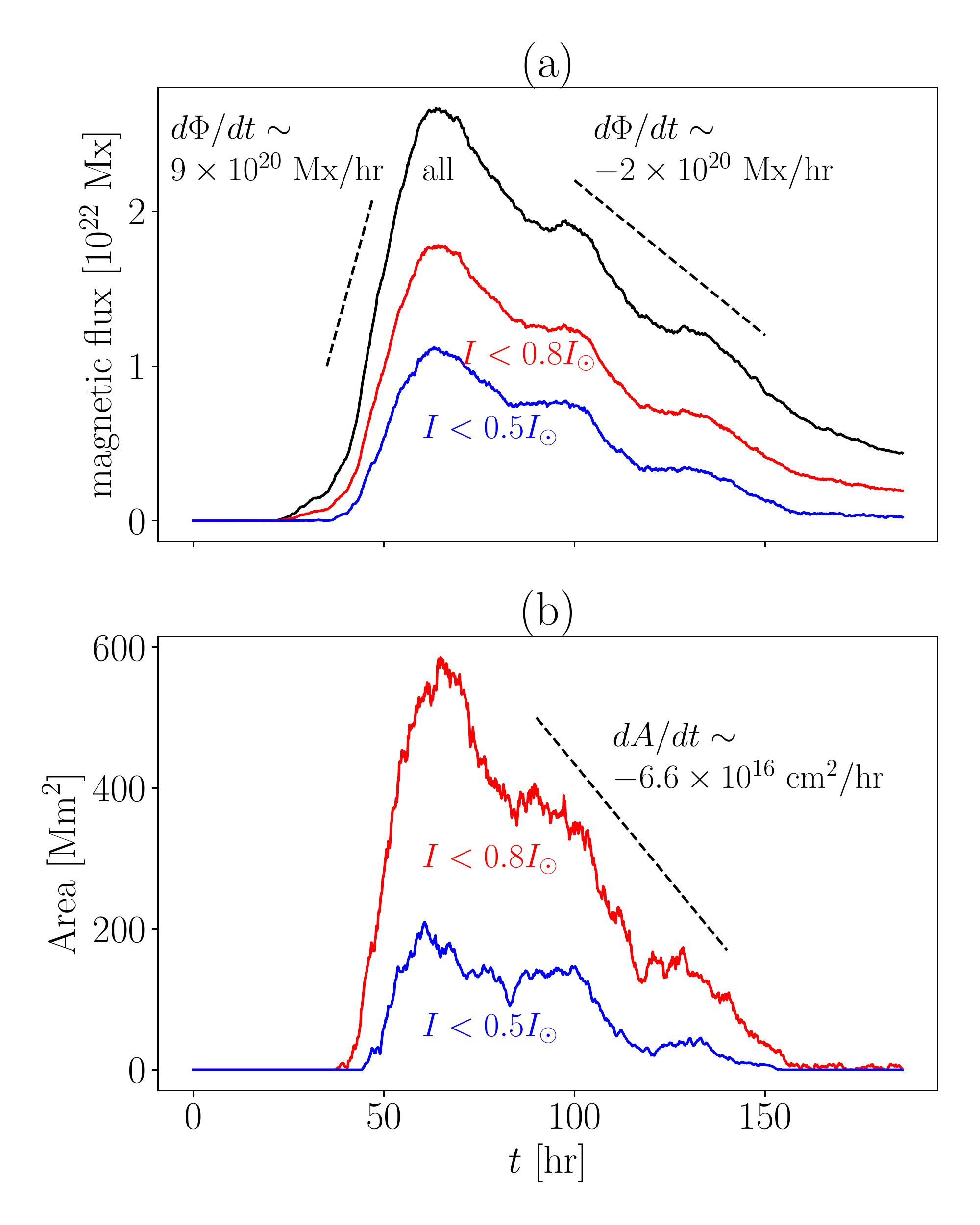}
  \caption{
    Temporal evolution of the unsigned magnetic flux (panel a) and sunspot area (panel b) at the $\tau=1$
      surface is shown. (a) The black line shows the unsigned
      magnetic flux of all areas. The red and blue lines
      show the unsigned magnetic flux at the area with
      $I<0.8I_\odot$ and $I<0.5I_\odot$, respectively, 
      where $I$ and 
      $I_\odot$ are the emergent intensity and that in 
      the quiet region, respectively. (b) The area of corresponding intensity is shown. The format of the lines is the same as in panel a.
  \label{magnetic_flux}
  }
\end{figure}

Fig. \ref{magnetic_flux} shows the temporal evolution of 
unsigned magnetic flux on
$\tau=1$ surface (panel a) and the area of the sunspot (panel b).
In Fig. \ref{magnetic_flux}a,
the black line shows the unsigned 
magnetic flux integrated over all horizontal domains. 
The red and blue lines show the unsigned magnetic flux with the
emergent intensity less than 80\% and 50\% of the averaged quiet sun
intensity $I_\odot$, respectively.
Fig. \ref{magnetic_flux}b shows the area with the emergent intensity less than $0.8I_\odot$ (red) and $0.5I_\odot$ (blue). For making Fig. \ref{magnetic_flux}b, we use the Gaussian filter with a width of 1 Mm.
The unsigned magnetic flux reaches its maximum of
$2.7\times10^{22}\ \mathrm{Mx}$
at $t=63.7$ hr.  We follow the definition of the emergence
time suggested by \cite{2013ApJ...762..130L}, in which the 
unsigned magnetic flux
reaches 10\% of the maximum magnetic flux.
The emergence time in this calculation is $t=37$ hr.
After the magnetic flux reaches its maximum, the magnetic flux begins
to decrease. Around $t=175\ \mathrm{hr}$, almost all the magnetic flux
in the sunspot $I<0.5I_\odot$ disappears, leading to approximate emergence and
decay rates
($d\Phi/dt$) of $9\times10^{20}\ \mathrm{Mx\ hr^{-1}}$ and 
$-2\times10^{20}\ \mathrm{Mx\ hr^{-1}}$, respectively. These values are similar 
to those in \cite{2014ApJ...785...90R}.
We note that they determine the rising speed of the flux tube at the bottom boundary,
while the combination of the convection and the magnetic field determines the rising property in this calculation.
\par
Compared with the observations, \cite{2011PASJ...63.1047O} and
\cite{2017ApJ...842....3N} show the relation of the maximum sunspot flux $\Phi_\mathrm{max}$
and the emergence rate $\left(d\Phi/dt\right)_\mathrm{e}$ as 
$\left(d\Phi/dt\right)_\mathrm{e}=9.6\times10^7\ \Phi_\mathrm{max}^{0.57}$ and 
$7.8\times10^{11}\ \Phi_\mathrm{max}^{0.36}$, where both are
in the unit of $\mathrm{Mx\ hr^{-1}}$. The maximum flux of the current simulation 
$\Phi_\mathrm{max}=2.7\times10^{22}\ \mathrm{Mx}$ leads to
$\left(d\Phi/dt\right)_\mathrm{e}=6\times10^{20}\ \mathrm{Mx}$ and $9\times10^{19}$ 
with
using \cite{2011PASJ...63.1047O} and \cite{2017ApJ...842....3N} relations, respectively.
In addition, \cite{2019ApJ...871..187N} show that the 95\% confidence interval of the emergence rate reaches $\left(d\Phi/dt\right)_\mathrm{e}=7\times10^{20}\ \mathrm{Mx\ hr^{-1}}$ with $\Phi_\mathrm{max}=2.7\times10^{22}\ \mathrm{Mx}$.
Compared with these observational results, this numerical simulation shows a slightly larger
emergence rate. 
\par
Regarding the decay rate of the sunspot, \cite{2008SoPh..250..269H} show 
the observational relation between the sunspot area $A_\mathrm{spot}$ and the dissipation rate
$dA_\mathrm{spot}/dt$ as $dA_\mathrm{spot}/dt = 
-3\times10^{16} - 5\times10^{-3} A_\mathrm{spot}\ [\mathrm{cm^2\ hr^{-1}}]$.
We assume that the mean magnetic field strength is 2000 G,
which is used in \cite{2019ApJ...871..187N}. This relation leads to
$d\Phi/dt = -1.9\times 10^{20}\ \mathrm{Mx\ hr^{-1}}$, which is consistent with the 
simulation in this study. Fig. \ref{magnetic_flux}b shows that the area of the sunspot reaches its maximum $A_\mathrm{spot}=5.9\times10^{18}\ \mathrm{cm^2}$ at $t=65$ hr and that the area decreases to zero at $t=154$ hr. This leads to the decay rate of $dA_\mathrm{spot}/dt=-6.6\times10^{16}\ \mathrm{cm^2\ s^{-1}}$. The \cite{2008SoPh..250..269H} relation leads to $dA_\mathrm{spot}/dt=-6.0\times10^{16}\ \mathrm{cm^2\ s^{-1}}$ for our case. Also, in terms of area, the calculation result is consistent with the observation.
\par
\subsection{Overall evolution}\label{overall_evolution}
\begin{figure}
  \centering
  \includegraphics[width=0.5\textwidth]{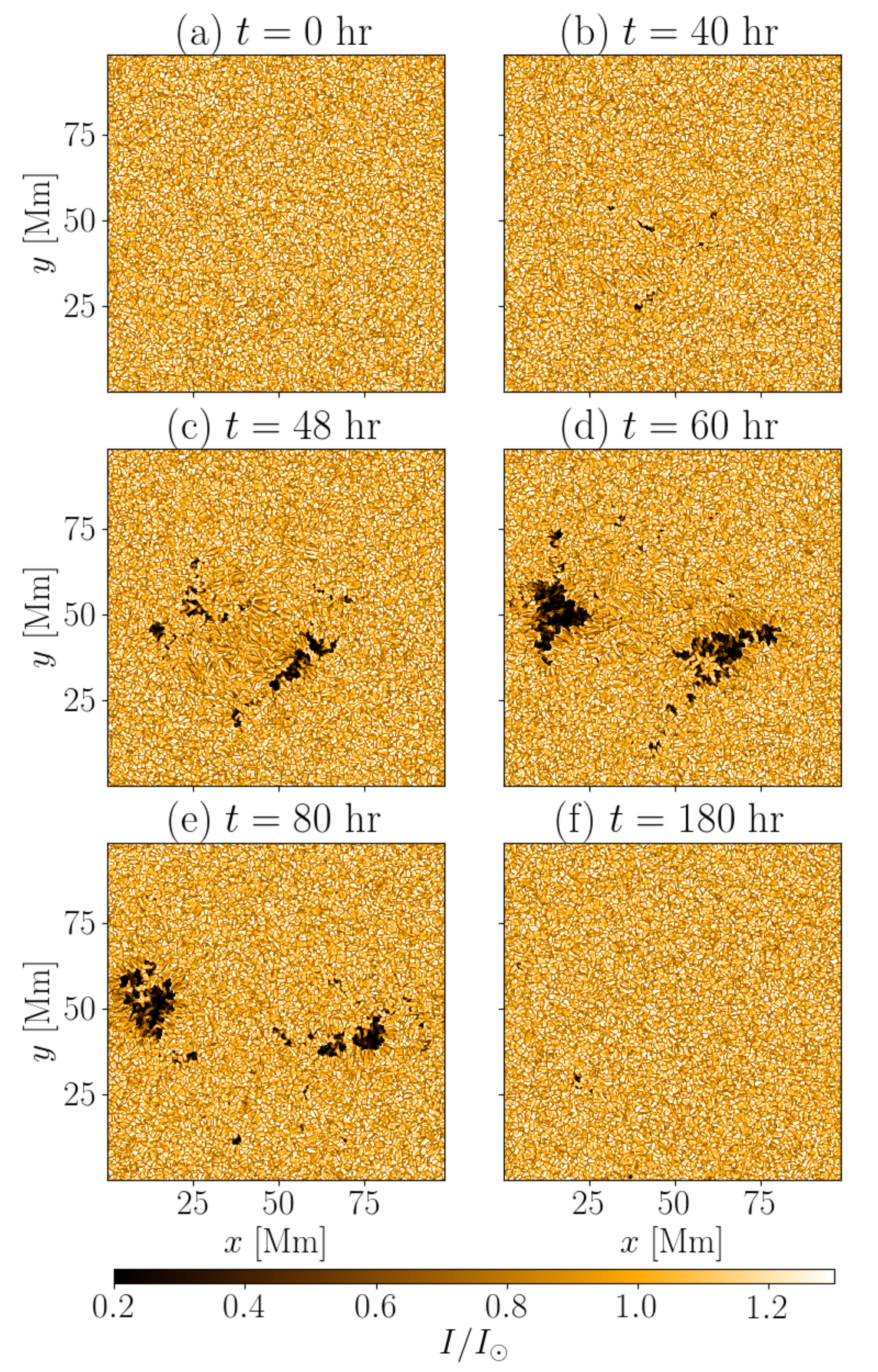}
  \caption{
    The temporal evolution of the emergent intensity is shown.
    The intensity is normalized with the mean quiet sun intensity $I_\odot$.
  \label{overall_in}
  }
\end{figure}

\begin{figure}
  \centering
  \includegraphics[width=0.5\textwidth]{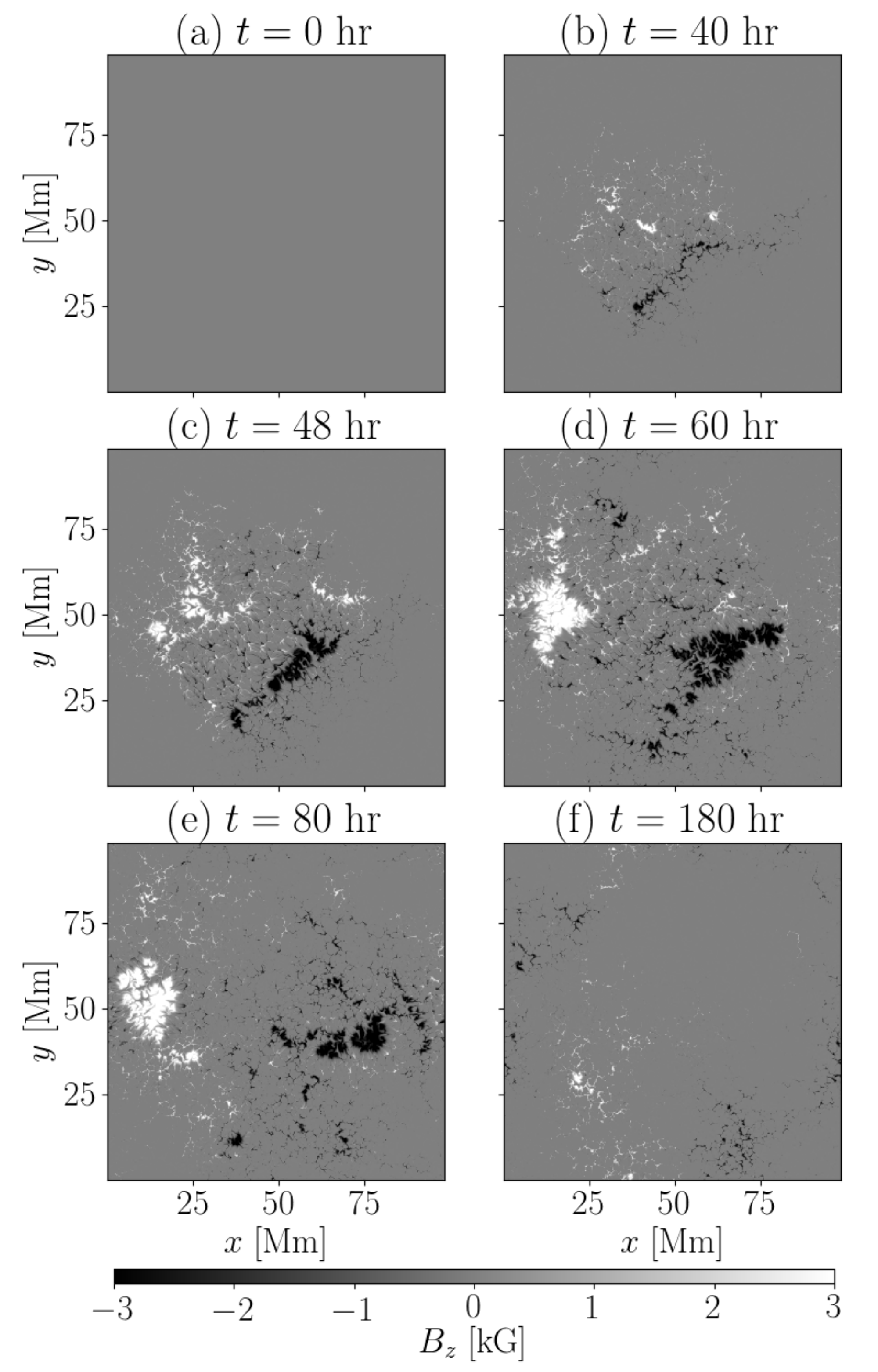}
  \caption{
    The temporal evolution of the line-of-sight magnetic field ($B_z$) at the
    $\tau=1$ surface is shown.
  \label{overall_bx}
  }
\end{figure}

Figs. \ref{overall_in} and \ref{overall_bx} show the emergent intensity and line-of-sight magnetic field $B_z$ at $\tau=1$ surface, 
respectively. Around the emergence time $t=37$ hr,
we begin to see
small pores in the intensity map 
(Fig. \ref{overall_in}b) and a diffused pattern in the magnetic field map
(Fig. \ref{overall_bx}b).
As time progresses, the small pores merge and construct the large-scale structure.
At $t=48\ \mathrm{hr}$, we observe elongated granules between the positive
and negative spots (Fig. \ref{overall_in}c and
\ref{overall_bx}c). When the photospheric magnetic flux reaches its 
maximum, a coherent sunspot appears (Fig. \ref{overall_in}d and \ref{overall_bx}d).
While we see some evidence of the penumbra around the sunspot,
the reproduced penumbra is much less prominent than the observation.
\cite{2012ApJ...750...62R} shows that the existence of the
penumbra is significantly affected by the top boundary condition.
Here, we use the potential magnetic field condition,
while \cite{2012ApJ...750...62R} argues that the horizontally
inclined magnetic field at the top boundary is required for a prominent
penumbra. We also note that \cite{2009Sci...325..171R} show
that the penumbra can be observed between two sunspots because the horizontal magnetic field is expected between them. In this study, the matured sunspots are far apart from each other, and even the 
penumbra between the sunspot pair cannot be observed.\par
After the magnetic flux reaches its maximum, the sunspots begin to
lose their magnetic flux. At $t=80\ \mathrm{hr}$, the right sunspot loses
significant flux, while the left sunspot keeps a coherent shape
with some bright features in the umbra (Fig. \ref{overall_bx}e).
At $t=180$ hr, almost all the magnetic flux disappears from the
umbra (Fig. \ref{overall_in}f and \ref{overall_bx}f).
\par
\begin{figure}
  \centering
  \includegraphics[width=0.5\textwidth]{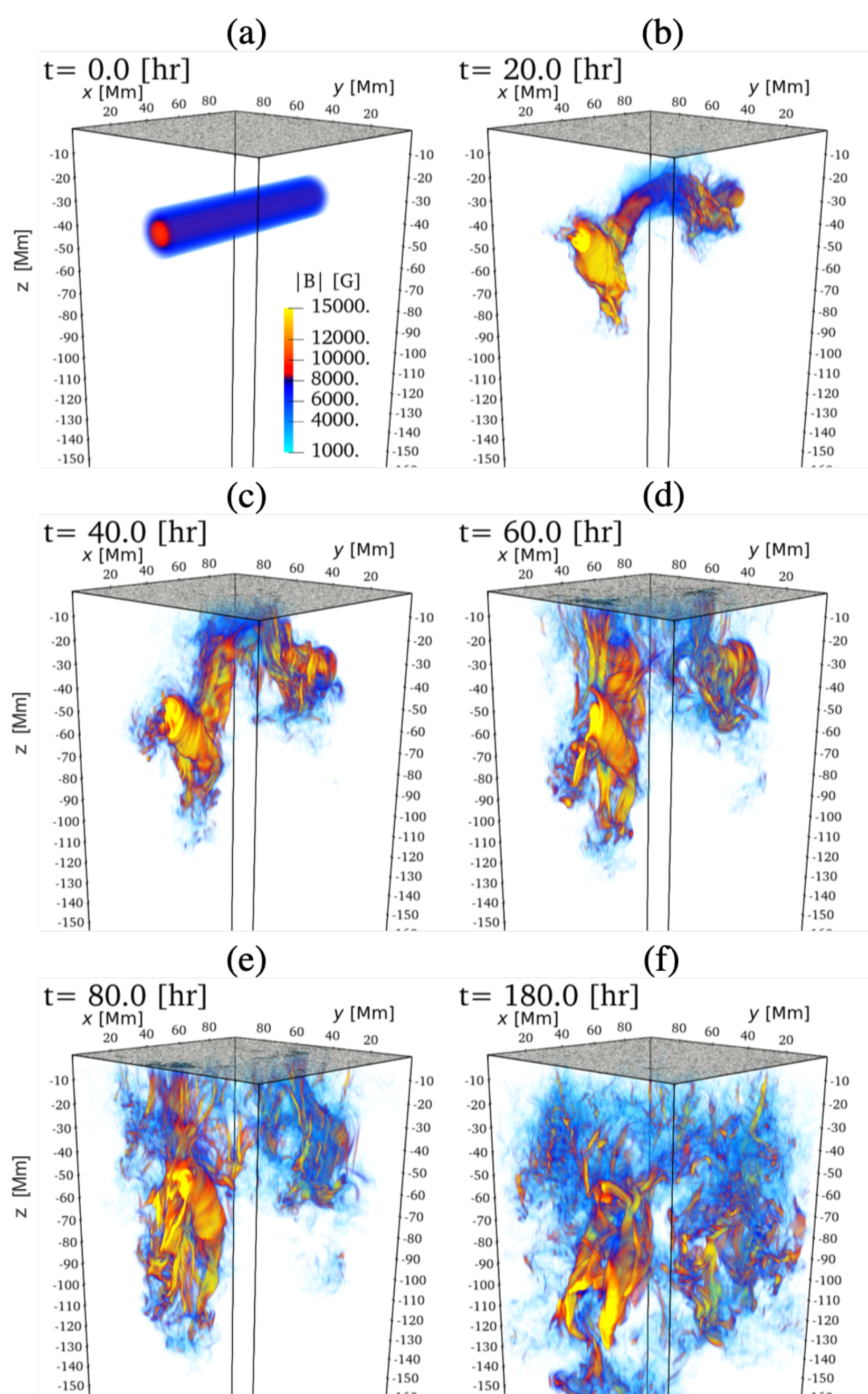}
  \caption{
    Temporal evolution of the three-dimensional structure of the
    overall magnetic field is shown. The volume rendering shows
    the magnetic field strength, and the gray surface around the top
    of the computational domain shows the emergent intensity.
  \label{3d}
  }
\end{figure}

Fig. \ref{3d} shows the three-dimensional structure of the magnetic field
deep in the convection zone. At $t=20\ \mathrm{hr}$, the magnetic field shows
an $\Omega$-shape structure with two anchoring downflows and a broad upflow
in the center region (Fig. \ref{3d}b). At $t=40\ \mathrm{hr}$ (Fig. \ref{3d}c), which is
around the emergence time, a significant fraction of the magnetic
flux reaches the near-surface layer ($z>-10\ \mathrm{Mm}$). At $t=60\ \mathrm{hr}$ (Fig. \ref{3d}d), which is around the time of the maximum photospheric magnetic flux, the root of the left sunspot reaches a depth of around 80 Mm, while the root of the right sunspot remains at a depth of around 30 Mm.
Below the photosphere, the magnetic field of the sunspot is mostly vertical.
At $t=80$ hr (Fig.\ref{3d}e), the sunspots begin to decay; at the same time, the subsurface structure
is destroyed. At $t=180$ hr (Fig. \ref{3d}f), when the sunspot disappears almost completely, most of the flux is transported downward, and coherent features can still be seen in the deep region below a depth of 60 Mm.
\par
\begin{figure}
\centering
\includegraphics[width=0.5\textwidth]{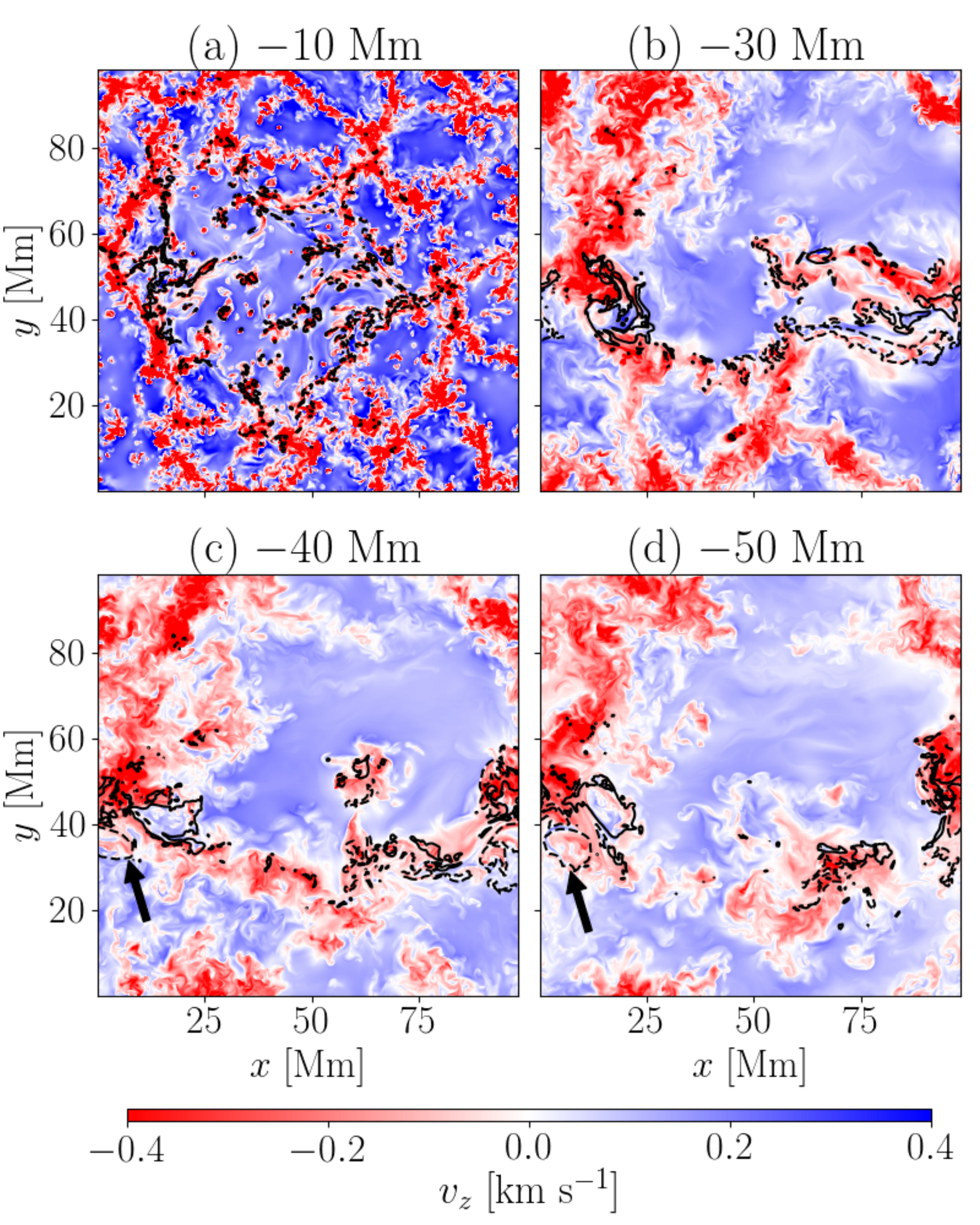}
\caption{
 The flow and magnetic structure in the deep layer at $t=60$ hr is shown.
  Panels a, b, c, and d show the layers at $z=-10$ Mm, $-30$ Mm, $-40$ Mm, and $-50$ Mm, respectively. The color contour shows the vertical velocity ($v_z$),
  and the black contour
  lines show the vertical magnetic field ($B_z$) of 5000 G (solid) and $-5000$ G (dashed).
\label{deep_bx}}
\end{figure}

The Coriolis force is an important factor for the tilt angle \citep{1991ApJ...375..761W} and the asymmetry of the sunspot pair \citep{1993A&A...272..621D}. \cite{2010ApJ...720..233C} show an almost symmetric sunspot pair because of their almost symmetric initial condition and setting. \cite{2014ApJ...785...90R} show an asymmetric sunspot pair resulting from the horizontal flow mimicking the Coriolis force-induced flow. \cite{2017ApJ...846..149C} also show asymmetry resulting from the boundary condition motivated by the dynamo calculation influenced by the Coriolis force. While the Coriolis force must be importanat to generate a statistical trend of the sunspot pair feature 
\citep{2011ApJ...741...11W}, our calculation without the Coriolis force can also cause some asymmetry in the sunspot pair. 
The asymmetry in this study reflects the convection flow structure in the deep convection zone.
The initial condition
(Fig. \ref{initial_condition}) shows that the left downflow has a more circular shape
than does the right downflow, which 
shows an elongated feature. Interestingly, the emerging sunspot follows this morphology. 
Fig. \ref{deep_bx} shows the flow
and magnetic structure in deep layers at $t=60$ hr. Color and line 
contours show the vertical velocity ($v_z$) 
and magnetic field ($B_z$), respectively. At a depth of at least 30 Mm (Fig. \ref{deep_bx}b),
we can still observe
the asymmetry of the magnetic feature,
i.e., the positive magnetic feature (solid line) shows the 
circular shape, while the negative feature (dashed line) is elongated.
In the deeper layer, the positive feature maintains a circular shape. 
The negative feature crosses the boundary and reaches the other side
indicated with a black arrow in Fig. \ref{deep_bx}c
because of the periodic boundary condition.
At a depth of 50 Mm, the negative feature
has
a circular shape (black arrow in Fig. \ref{deep_bx}d).
We note that the time scale of the convection at 50 Mm depth is about 5 days, and the structure of the magnetic flux tube has not been distorted well at this depth.
This result indicates that the shape of the sunspot is influenced by the deep convection structure at least 30 Mm below the photosphere. 
\par

\subsection{Rising mechanism of flux tube} \label{rising_mechanism}
\begin{figure}
  \centering
  \includegraphics[width=0.5\textwidth]{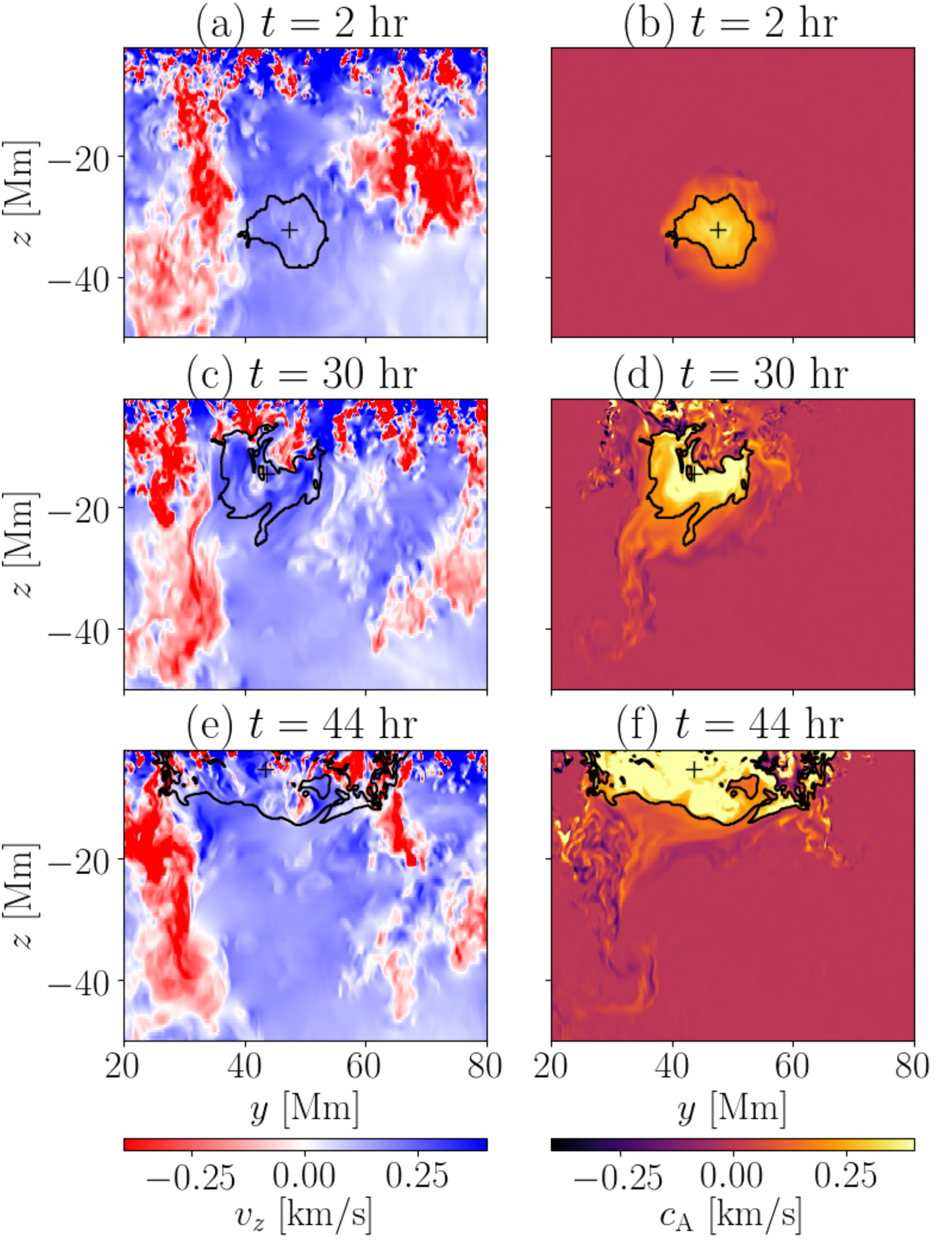}
  \caption{Rising process at $x=41\ \mathrm{Mm}$. The left and right panels
  show the vertical velocity $v_z$ and Alfven velocity of the axial field
  $c_\mathrm{A}=B_x/\sqrt{4\pi\rho}$, respectively. The top, middle, and bottom panels 
  show results at $t=2$, $30$, and $44$ hr, respectively. The black lines show the 
  boundary of the detected flux tube. The black cross shows the center of the flux tube.}
  \label{rising_2d}
\end{figure}

\begin{figure}
  \centering
  \includegraphics[width=0.5\textwidth]{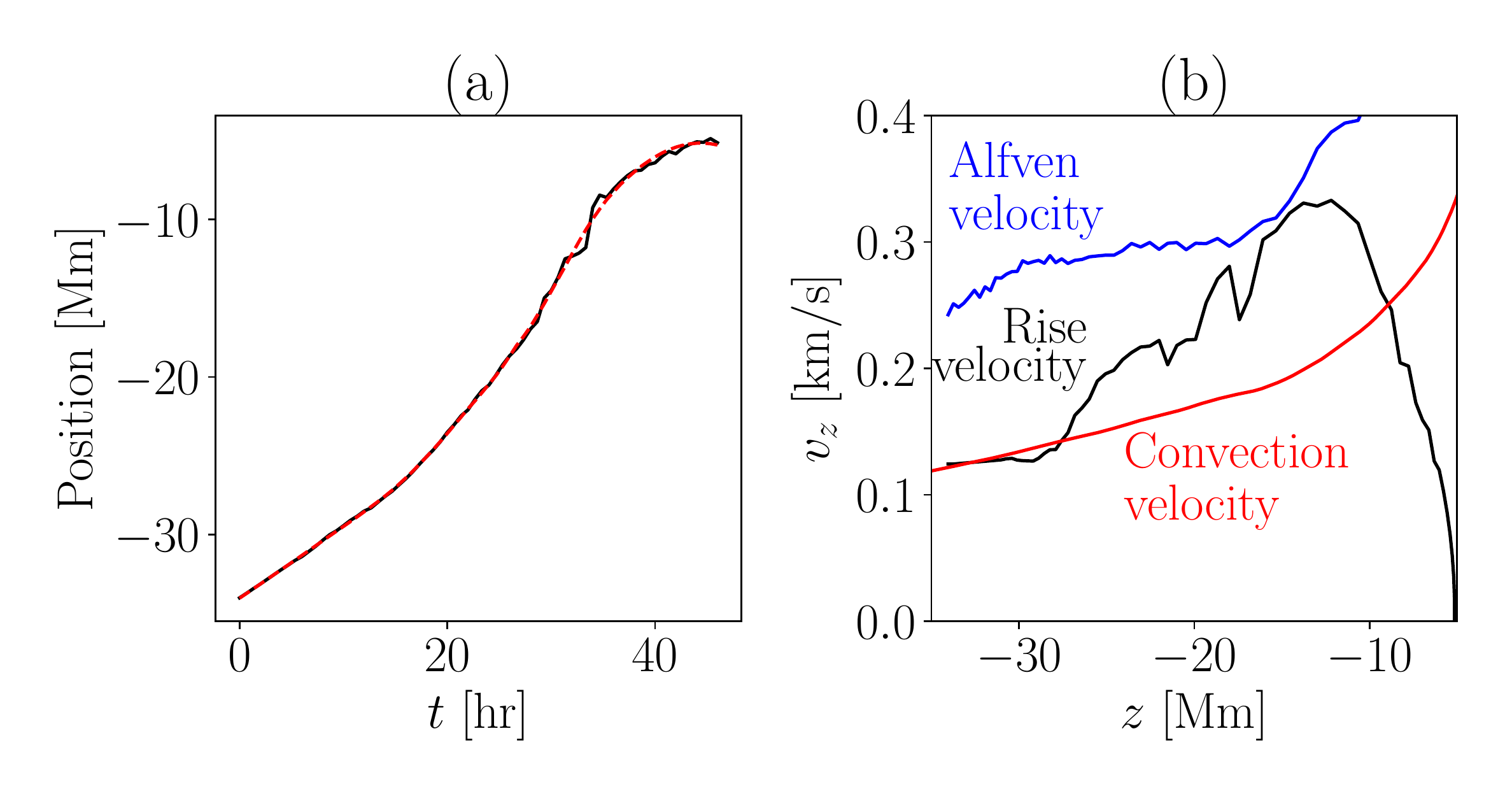}
  \caption{(a) The time evolution of the center of the flux tube in height 
  $z_\mathrm{c}$
  is shown. The black line is the raw data, and the red 
  line is filtered data for estimating the rising velocity. (b) The black
  line shows the rising speed of the center of the flux tube. The red line shows 
  the RMS convection velocity in the calculation without
  the magnetic field.
  The blue line shows the local Alfven velocity 
  $c_\mathrm{A}$
  averaged in the flux 
  tube ($S_\mathrm{c}$).}
  \label{flux_tube_rising}
\end{figure}

In this section, we investigate the rising process of the flux tube mainly
around the center region ($x=41$ Mm). Fig. \ref{rising_2d} shows the temporal
evolution of the rising flux tube. The left and right panels show the vertical 
velocity $v_z$ and Alfven velocity of the axial field
$c_\mathrm{A}=B_x/\sqrt{4\pi\rho}$, respectively. To investigate the motion of
the flux tube, we adopt the following definition of the center of the flux tube.
Basically, we use a clumping method.
We detect clumps of the region having an Alfven velocity beyond a given threshold.
With iteration, we survey the threshold $c_\mathrm{A0}$ with which the largest clump has 
a magnetic flux of $6\times10^{21}\ \mathrm{Mx}$, where the magnetic flux is
estimated with $B_x$. The largest clump ($S_\mathrm{c}$) in each time step 
is defined as the rising flux tube. The center of the flux tube
$(y_\mathrm{c},z_\mathrm{c})$ is defined as follows:
\begin{eqnarray}
  y_\mathrm{c} &=& \frac{\int_{S_\mathrm{c}}yc_\mathrm{A} dS}{\int_{S_\mathrm{c}}c_\mathrm{A} dS}, \\
  z_\mathrm{c} &=& \frac{\int_{S_\mathrm{c}}zc_\mathrm{A} dS}{\int_{S_\mathrm{c}}c_\mathrm{A} dS}.
\end{eqnarray}
The boundary and center of the flux tube are shown with black lines and black crosses
in Fig. \ref{rising_2d}, respectively. The time evolution of the center of the flux tube is
shown in Fig. \ref{flux_tube_rising}. Fig. \ref{flux_tube_rising}a shows the vertical position
of the center of flux tube $z_\mathrm{c}$. In the beginning, the rising speed increases 
with time, when the flux tube reaches the solar surface, the speed is reduced. The time profile
of the vertical position is not smooth because of sudden merging and splitting. To
evaluate the rising speed, we use a Savitzky--Golay filter for the vertical position,
which is shown as the red line in Fig. \ref{flux_tube_rising}a. By using the filtered
profile of the vertical position, we estimate the rising speed of the center of the
flux tube. The black line in Fig. \ref{flux_tube_rising}b shows the rising speed
of the flux tube. The 
red line shows the root-mean-square (RMS) upflow convection velocity
in a hydrodynamic run, i.e., without the magnetic field. We note that
the downflow convection velocity is typically larger than the upflow velocity.
The blue line shows the mean Alfven velocity in the flux tube.
In the beginning, the rising speed is almost the same as the upward convection velocity.
This is natural because the initial flux tube is force-free and the flux tube needs to obey
the convective flow. As time progresses, the rising speed exceeds the convection velocity, indicating some contribution of the magnetic field to the rising process. 
Even at the
end of the rising process, the rising speed does not reach the local Alfven velocity.
This result indicates that the rising process is neither a pure magnetic nor a pure convective 
process. \par
\begin{figure}
  \centering
  \includegraphics[width=0.5\textwidth]{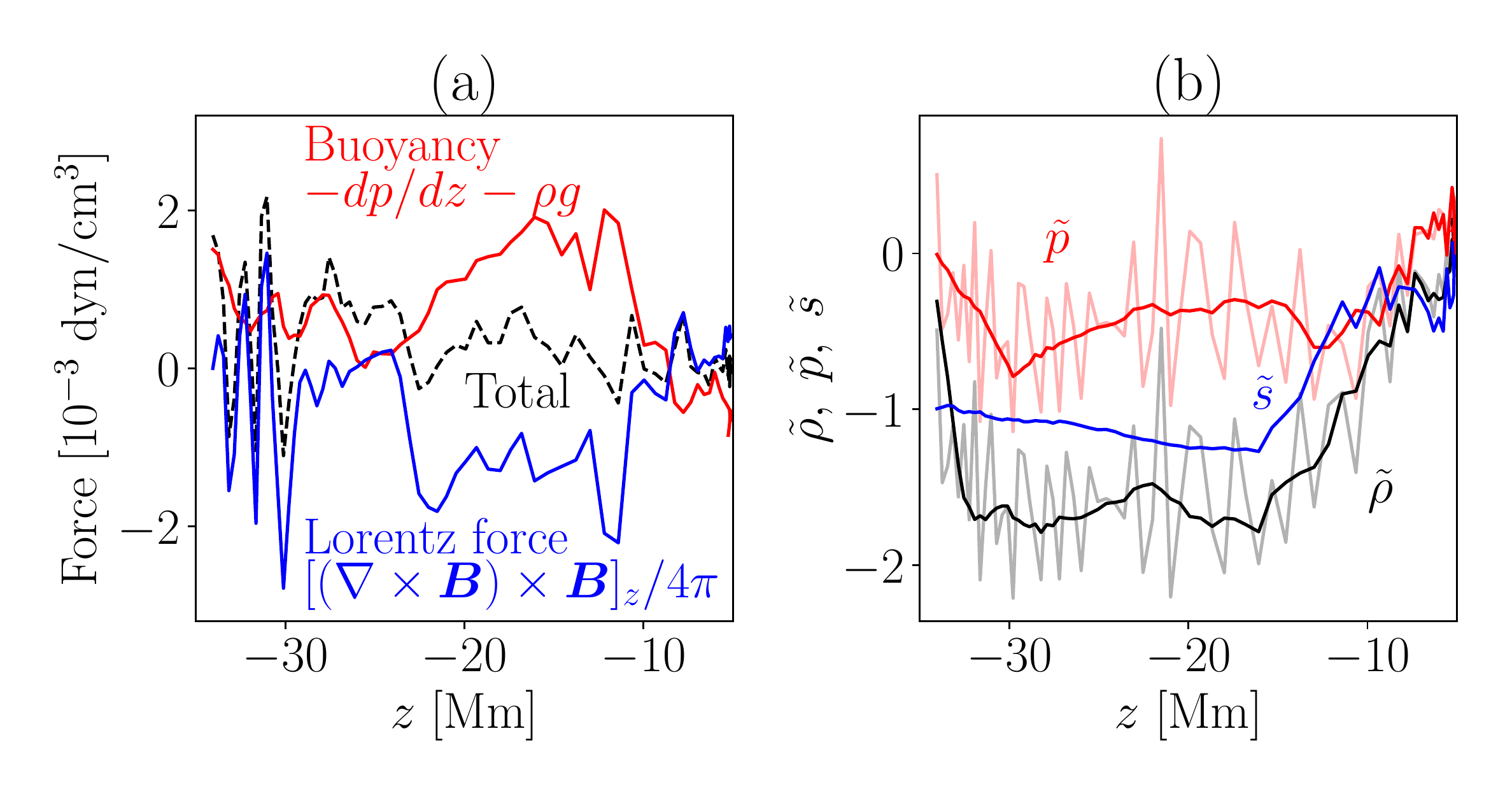}
  \caption{(a) The mean forces in the flux tube are shown.
  The red and blue lines show the buoyancy and Lorentz force, respectively.
  The black line shows the total force, including the inertia.
  (b) The mean normalized density, pressure, and entropy
  in the flux tube are shown.
  The light color in the panel b shows the values before 
  filtering out the sound waves.}
  \label{flux_tube_eom}
\end{figure}

To understand the rising mechanism, we investigate the
vertical equation of motion.
Fig. \ref{flux_tube_eom}a shows the vertical forces averaged in the flux tube.
Before evaluating the equation of motion, we filtered out the sound waves in the phase space 
\citep{2007ApJ...657.1157G}.
The equation of motion in the vertical direction is the following:
\begin{eqnarray}
  \rho \frac{\partial v_z}{\partial t} = -\rho \left(\bs{v}\cdot \nabla\right)v_z
  - \frac{\partial p}{\partial z} - \rho g + \frac{1}{4\pi}
  \left[
  \left(\nabla\times\bs{B}\right)\times\bs{B}\right]_z.
\end{eqnarray}
The red and blue lines in Fig. \ref{flux_tube_eom}a show the buoyancy 
($-dp/dz - \rho g$) and Lorentz force ($[(\nabla\times\bs{B}\times\bs{B})]_z/4\pi$), respectively.
The moving flux tube can be roughly regarded as a Lagrangean parcel in the $y-z$ plane.
Only the contribution from the inertia term ($-\rho \left(\bs{v}\cdot \nabla\right)v_z$)
to the motion of the rising flux tube is ($-\rho v_x dv_z/dx$).
The black line is the sum of these terms.
The buoyancy and total force direct upward during the rising process, while the Lorentz force
directs downward. The main contribution of the Lorentz force is magnetic tension.
The roots of the magnetic flux are anchored in the deep layer; thus, the magnetic tension in the rising part directs downward. 
Fig. \ref{flux_tube_eom}b shows the mean normalized density, pressure,
and entropy in the
flux tube. The normalization procedure, i.e., the tilde, 
for quantities are defined as follows:
\begin{eqnarray}
  \tilde{\rho} &=& \frac{\rho - \langle \rho\rangle_\mathrm{hd}}{\rho_\mathrm{rms(hd)}},
  \label{norm_rho}\\
  \tilde{p} &=& \frac{p - \langle p\rangle_\mathrm{hd}}{\rho_\mathrm{rms(hd)}}
  \left(\frac{\partial \rho}{\partial p}\right)_s
  ,\label{norm_p}\\
  \tilde{s} &=& \frac{s - \langle s\rangle_\mathrm{hd}}{\rho_\mathrm{rms(hd)}} 
  \left(\frac{\partial \rho}{\partial s}\right)_p\label{norm_s}
  , 
\end{eqnarray}
where $\langle q\rangle_\mathrm{hd}$ and $q_\mathrm{rms(hd)}$ are the horizontal average and RMS
value of a quantity $q$ in the calculation without the magnetic field, respectively. 
After normalization, the density, pressure, and entropy are related as
$\tilde{\rho}=\tilde{s}+\tilde{p}$. We also note that because $(\partial \rho/\partial s)_p$ is negative, the negative value of $\tilde{s}$ corresponds to a positive perturbation of the entropy $s_1 - \langle s_1 \rangle_\mathrm{hd}$ and vice versa.
The result shows that the
mean normalized density in the flux tube is significantly low (the perturbation is 1.7 times 
larger than the RMS value); this is the primary driver of the rising flux tube.
There are two contributions to the low density in the flux tube. The first is the
low pressure caused by the magnetic pressure. After the rising starts, the twist becomes
relaxed and the force-free balance is destroyed. Then, the gas pressure gradient force 
plays a role in maintaining the flux tube and decreasing the gas pressure in the flux tube.
Because the initial flux tube is force-free, the magnetic tension plays 
a role in balancing the magnetic pressure,
even during the rising process; thus, the gas pressure is not significantly low in this case.
The other contribution to the low density is the high entropy maintained by the suppression of the mixing.
The initial magnetic flux tube is located in the upflow region and has higher entropy than does the surrounding plasma.
As seen in Fig. \ref{rising_2d}c and d, the magnetic field suppresses
the mixing between upflows and downflows. In an ordinary medium in the convection zone,
the upflow warm medium is mixed in a mixing length and loses its high entropy, but
this magnetic flux tube can avoid this process and maintain high entropy, 
leading to continuously low density in the flux tube and further acceleration. 
Fig. \ref{flux_tube_eom}b shows that the high entropy is the main contribution to the low
density in the flux tube.

\subsection{Surface flow prior to emergence time}
In this section, we investigate the divergent flow prior 
to emergence time to
compare the results with \cite{2016SciA....2E0557B},
in which clear divergent flow is not seen at 3 hr before 
the flux emergence.\par
\begin{figure}
  \centering
  \includegraphics[width=0.3\textwidth]{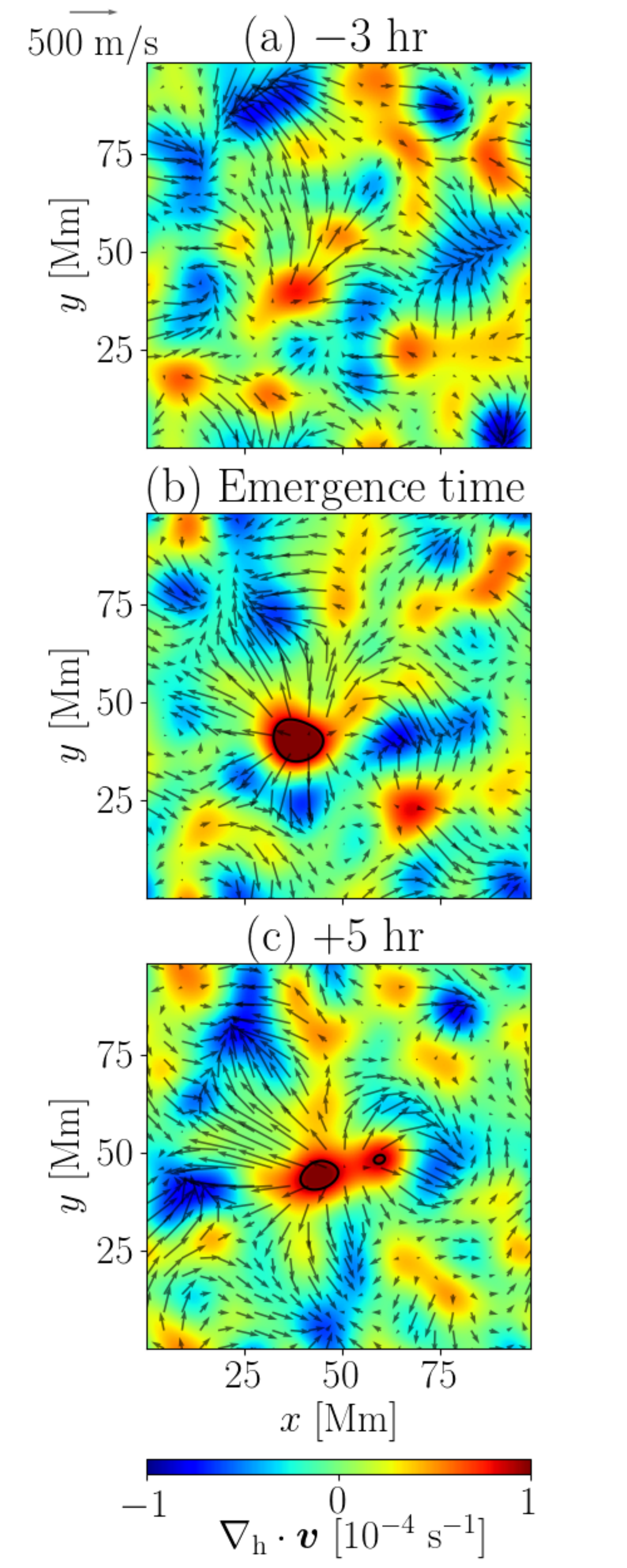}  
  \caption{The divergence of the horizontal flow at $\tau=1$
  surface. Panel b shows the result at the emergence time 
  ($t=37$ hr). Panels a and c show 
  3 hr before and 5 hr after
  the emergence time, respectively. The arrows show the horizontal flow. The
  black contour line shows the 3$\sigma$ value of the horizontal divergence 
  ($\nabla_\mathrm{h}\cdot\bs{v}=9\times10^{-9}\ \mathrm{s^{-1}}$)}
  \label{div_flow}
\end{figure}

 
To this end, we follow a similar procedure to that of \cite{2016SciA....2E0557B}.
We take the divergence of the horizontal flow at $\tau=1$ surface and 
the Gaussian filter with a 6-Mm window. Fig. \ref{div_flow} shows the 
horizontal divergence $\nabla_\mathrm{h}\cdot\boldsymbol{v}$ at $\tau=1$ surface at $t=34$ and 37, and 42 hr. 
The standard error of the distribution is 
$3\times10^{-5}\ \mathrm{s^{-1}}$, and the black line in Fig. \ref{div_flow}
shows the 3$\sigma$ value. 
We again
note that $t=37$ hr is the emergence time. Panels a and c are
5 hr after and 3 hr before the emergence time. While we see a clear
divergent flow at and after the emergence time 
(panels b and c), we do
not see it 3 hr before
the emergence time (panel a).
This result is consistent with the observation of
\cite{2016SciA....2E0557B}. By contrast, the rising speed of the
flux tube exceeds $250\ \mathrm{m\ s^{-1}}$ at a depth of 18 Mm,
while \cite{2016SciA....2E0557B} suggest that the rising speed is
no larger than $150\ \mathrm{m\ s^{-1}}$ to avoid 
divergent flow prior to the emergence time.\par
\begin{figure}
  \centering
  \includegraphics[width=0.5\textwidth]{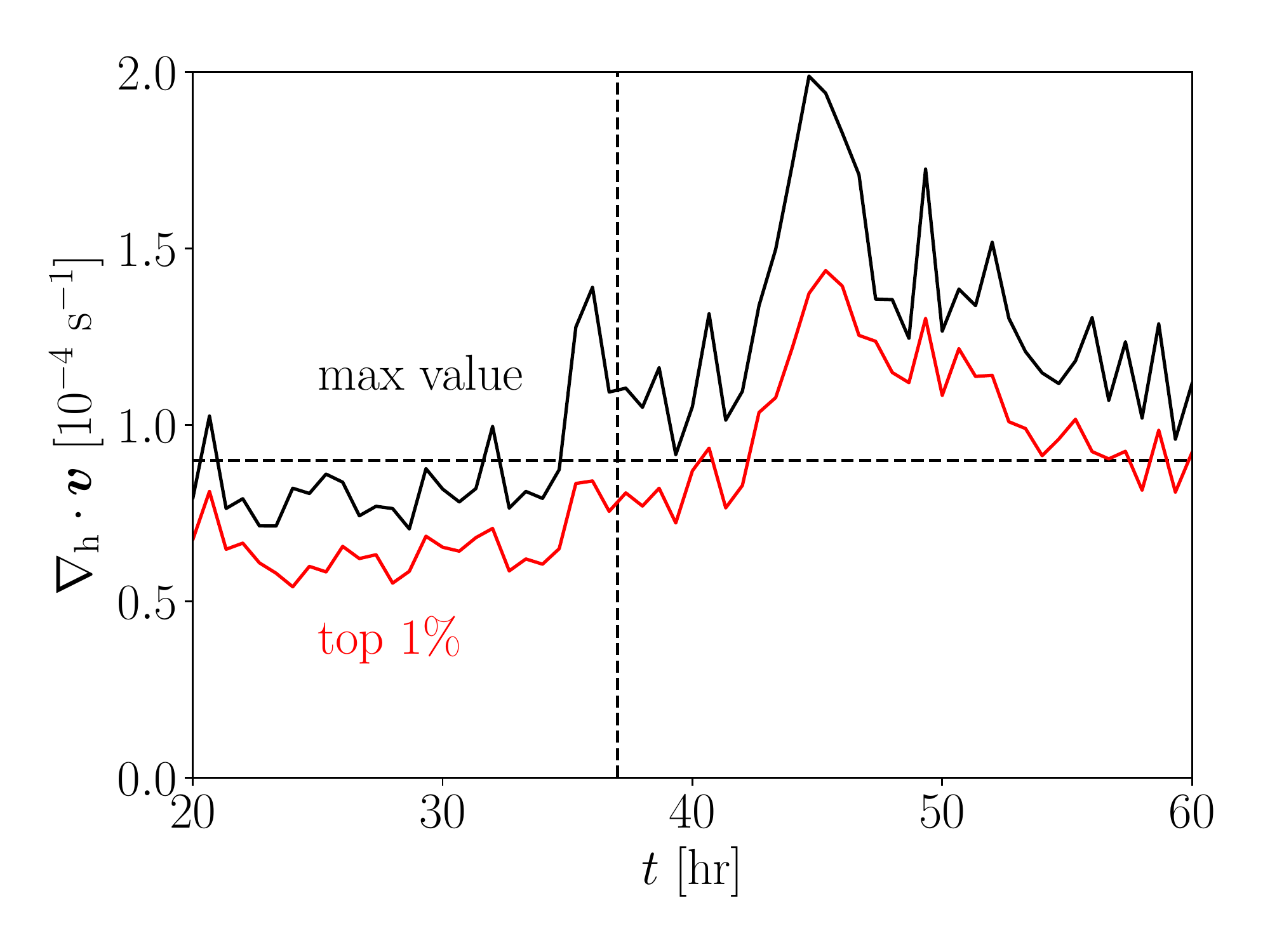}
  \caption{Temporal evolution of the horizontal divergence at $\tau=1$ surface.
  The black and red lines show the maximum and top 1\% value of the 
  divergence, respectively. The horizontal and vertical dashed lines show 
  3$\sigma$ value and the emergence time, respectively.}
  \label{div_flow_time}
\end{figure}

 
Fig. \ref{div_flow_time} shows the temporal evolution of the 
horizontal divergence at $\tau=1$ surface.
The black and red lines show the maximum and 
top 1\% value of the divergent flow, respectively. This also indicates that the 
clear evidence of the divergent flow is seen at only 1 hr before 
the emergence time.

\subsection{Flow in flux tube}\label{flow_tube}
\begin{figure}
  \centering
    \includegraphics[width=0.5\textwidth]{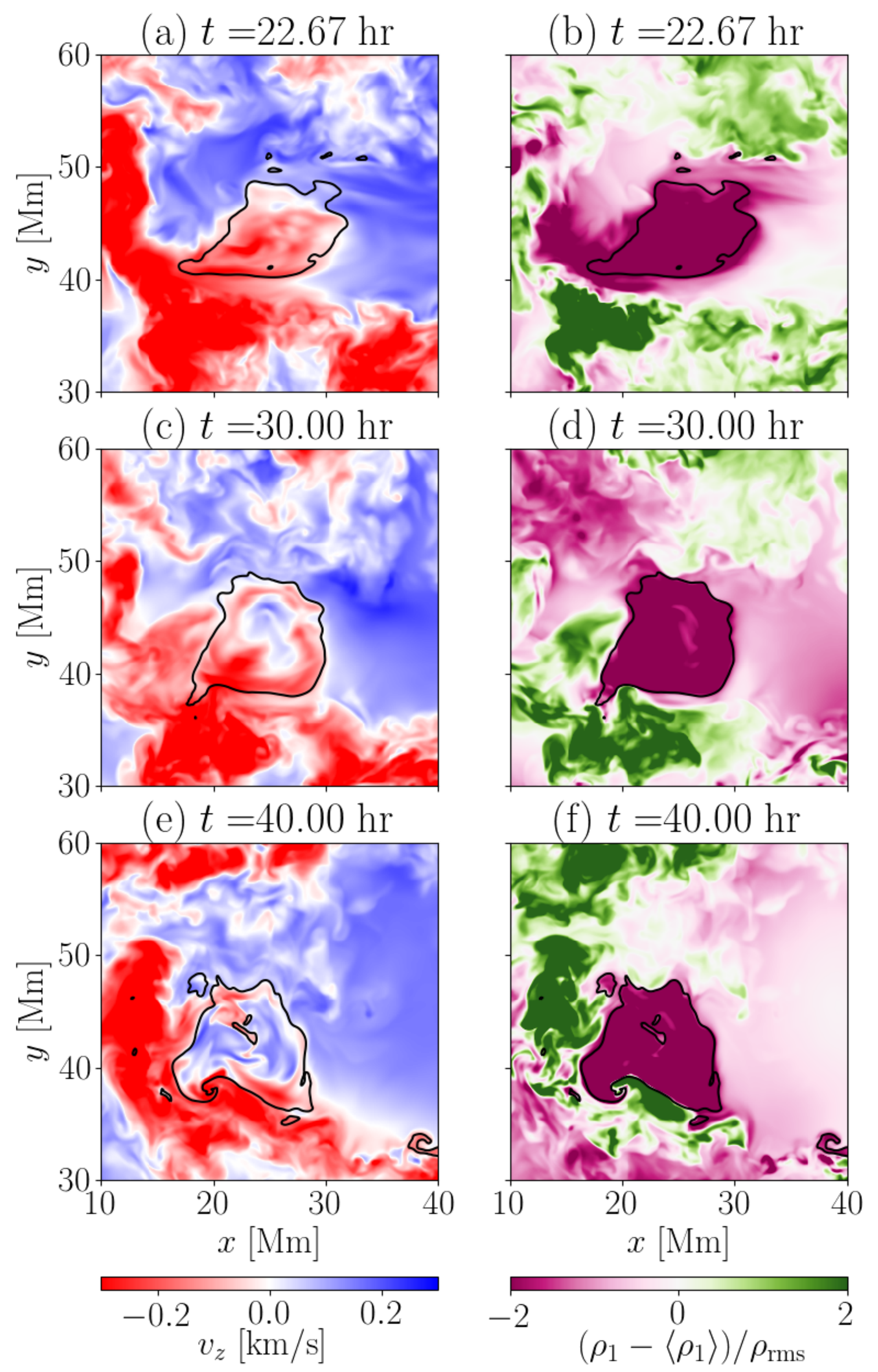}
  \caption{The left and right panels show the vertical flow and normalized density 
  at $z=-30\ \mathrm{Mm}$, respectively.
  The top, middle, and bottom panels show the results at $t=22.67,\ 30$, and $40$
  hr, respectively. The black contour lines show vertical magnetic field $B_z$
  of 7000 G, roughly indicating the magnetic flux tube.}
  \label{tube_internal_flow}
\end{figure}

In this section, we investigate the flow in the flux tube during
the flux emergence process. First, we discuss the vertical
flow at $z=-30$ Mm in detail. Then, we discuss the 
flow at different heights. Fig. \ref{tube_internal_flow} shows the 
temporal evolution of the vertical flow 
and the density in the flux tube. 
From this section, we focus on the left coherent positive sunspot.
Because the root of the sunspot is created with the downflow
from the horizontal magnetic flux tube, the root is initially
filled with the downflow (Fig. \ref{tube_internal_flow}a: 
$t=22.67$ hr). This is somewhat consistent with
\cite{2017ApJ...846..149C}, in which the strong magnetic 
concentration occurs at a coherent downflow in the deep 
layer. A difference is seen in the following evolution.
The flow in the flux tube changes signs around $t=30$ hr in 
Fig. \ref{tube_internal_flow}c.
After the sunspot is formed (Fig. \ref{tube_internal_flow}e: $t=40$ hr),
the flux tube is filled with the upflow.
During this process, the flux tube is filled with the low-density medium
(Figs. \ref{tube_internal_flow}b, d, and f).
\par
\begin{figure}
  \centering
  \includegraphics[width=0.3\textwidth]{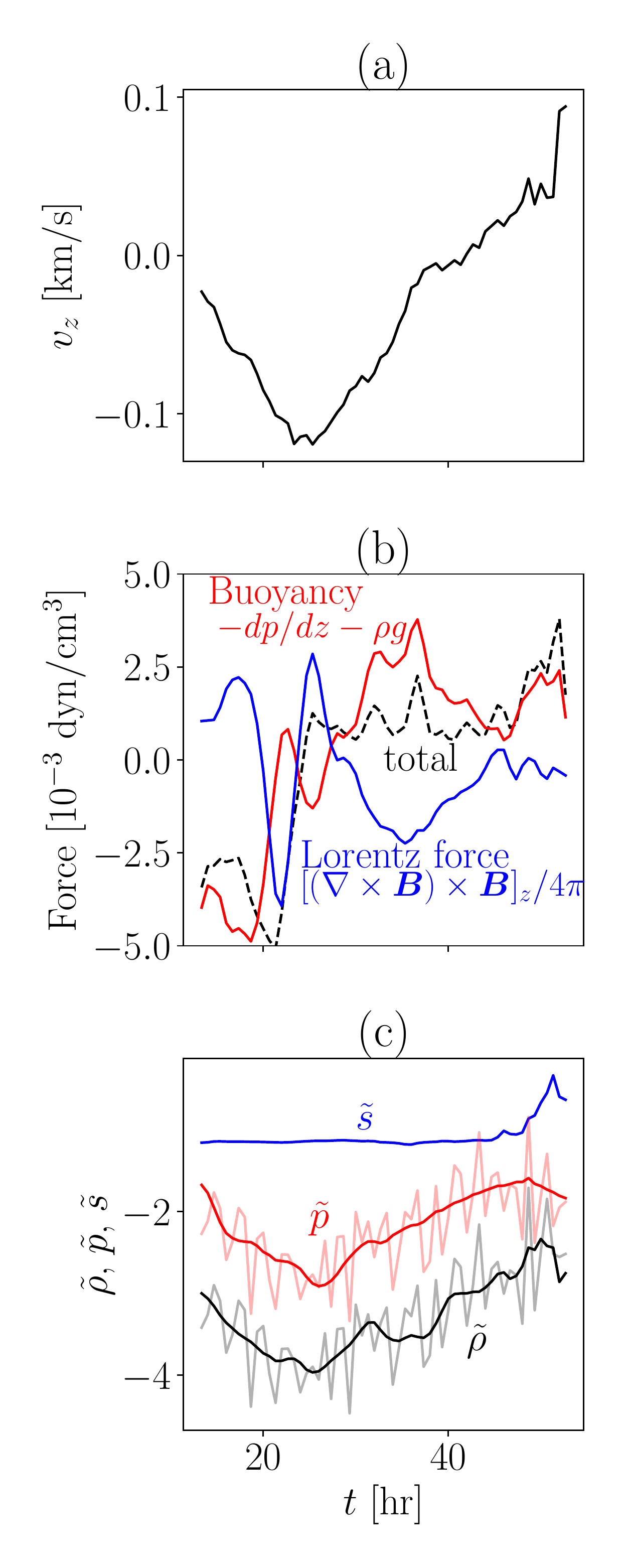}
  \caption{(a) Temporal evolution of the mean vertical velocity
  in the flux tube at a depth of $30$-Mm is shown. (b) The force
  balance in the flux tube at a depth of $30$ Mm is shown. The red, blue,
  and black lines show the buoyancy, Lorentz, and total forces,
  respectively. (c) The black, red, and blue lines show the normalized
  density, pressure, and entropy, respectively. 
  The light color in panel c shows the value before filtering out the sound waves.}
  \label{eom_tube_internal}
\end{figure}

Here, we define the flux tube as the largest clump in which the
vertical magnetic field exceeds the threshold value of 7000 G.
Fig. \ref{eom_tube_internal}a shows the temporal evolution of the
mean vertical flow in the flux tube.
In the beginning, the flux tube is filled with the downflow.
Around $t=23$ hr, the downflow stops increasing in amplitude.
Around $t=40$ hr, the vertical flow changes signs.
Afterward, the amplitude of the upflow increases continuously.
Fig. \ref{eom_tube_internal}b shows the force balance in the flux
tube. The red and blue lines show the buoyancy and the Lorentz force, respectively.
The flux tube roughly acts as the Lagrangean parcel in the $x-y$ plane.
Thus, for the total dynamics, the contribution from the inertia term is
the vertical inertia term $-\rho v_z\partial v_z/\partial z$.
The black line in Fig. \ref{eom_tube_internal}b shows the sum of the buoyancy,
Lorentz force, and vertical inertia terms. The total force is
consistent with the temporal evolution of the vertical velocity in the
flux tube. Around $t=23$ hr, the total force changes its sign and keeps
the positive value after that. The main driver of the upflow in the
flux tube is the buoyancy.
Fig. \ref{eom_tube_internal}c shows the normalized density, pressure, and entropy averaged in
the flux tube (see eqs. (\ref{norm_rho})--(\ref{norm_s})).
In this case, the main contribution of the low density is the low gas pressure 
caused by the adiabatic expansion from the Lorentz force.
\par
\begin{figure}
  \centering
  \includegraphics[width=0.5\textwidth]{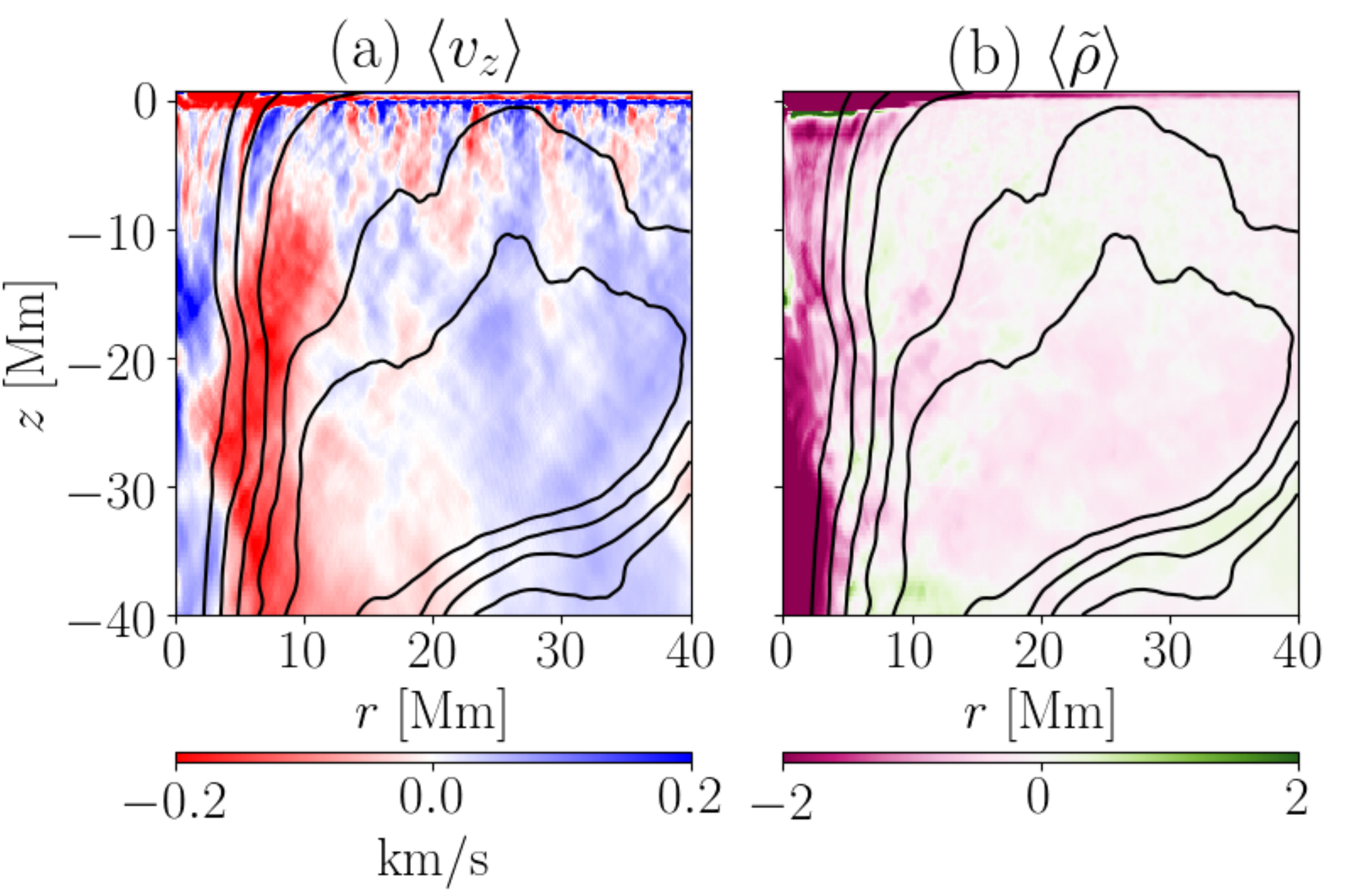}
  \caption{Azimuthally averaged values in the polar coordinate at
  $t=66$ hr
  are shown. (a) The vertical velocity ($v_z$) and (b) normalized
  density ($\tilde{\rho}$) are shown. The black contour lines show the
  magnetic flux evaluated from the center of the coordinate 
  ($r=0$).
  The values are from $2\times10^{21}\ \mathrm{Mx}$ to 
  $1\times10^{22}\ \mathrm{Mx}$.}
  \label{deep_vertical_flow}
\end{figure}

Fig. \ref{deep_vertical_flow} shows the azimuthally averaged values in the polar
coordinate. Because the generated sunspot and its flux tube are tangling
along the vertical direction,
different heights cannot share the same origin as the polar coordinate.
Thus, we define the origin of the polar coordinate at every depth
separately. We adopt the following procedure.
\begin{enumerate}
  \item We estimate the magnetic flux of the positive vertical magnetic field 
        in the left half of the computational domain ($\Phi_+$).
  \item The threshold value of the magnetic flux ($\Phi_\mathrm{s}$)
        is $25\%$ of $\Phi_+$.
  \item We search the critical value $B_\mathrm{c}$ for the vertical
        magnetic field $B_z$ with which the largest clump
        has a vertical magnetic flux of $\Phi_\mathrm{s}$, where the definition of
        the clump is the continuous region in which the vertical magnetic field
        exceeds the critical value. The flux tube is identified with $S_\mathrm{o}$.
  \item The origin of the polar coordinate $(x_\mathrm{o},y_\mathrm{o})$ at depth 
  is defined with the following:
  \begin{eqnarray}
    x_\mathrm{o} &=& \frac{\int_{S_\mathrm{o}}xB_zdS}{\int_{S_\mathrm{o}}B_zdS}, \\
    y_\mathrm{o} &=& \frac{\int_{S_\mathrm{o}}yB_zdS}{\int_{S_\mathrm{o}}B_zdS}.  
  \end{eqnarray}
  \item If $\Phi_+$ is smaller than $2.5\times 10^{21}\ \mathrm{Mx}$, we use the origin
        of the polar coordinate defined at a place one grid below.
  \item The vertical profile of the origin of the polar coordinate is not smooth. We obtain
        the smooth profile of the origin of the polar coordinate with the Savitzky--  Golay filter.
        The filtered profile is used for the origin of the polar coordinate in the following analyses.
\end{enumerate}
Fig. \ref{deep_vertical_flow} shows the azimuthally averaged profile of the vertical velocity
(Fig. \ref{deep_vertical_flow}a) and the normalized density $\tilde{\rho}$
(Fig. \ref{deep_vertical_flow}b) at $t=66$ hr, where the parenthesis $\langle\rangle$
shows the azimuthal average at constant radius $r$.
The finding at a depth of 30 Mm is applicable to different depths. Except for the near-surface layer ($z> -10$ Mm),
the center of the flux tube is filled with the upflow, surrounded by the coherent downflow in the
outer side of the flux tube. The coherent upflow corresponds to low density
(Fig. \ref{deep_vertical_flow}b), indicating that the driving mechanism of the upflow is
the buoyancy.

\subsection{Formation mechanism of sunspot}

In this section, we discuss the formation mechanism of sunspots.
The most important process for the formation of sunspots is the 
collection of the magnetic flux toward the center region. As discussed in 
\cite{2010ApJ...720..233C}, the temporal evolution of the magnetic
flux is expressed as follows:
\begin{eqnarray}
  \frac{\partial\Phi_\mathrm{c}}{\partial t} &=&
  \frac{\partial}{\partial t}\int_S 
  B_z dS \nonumber\\
  &=&
  2\pi r\left(
     \langle v_z\rangle \langle B_r\rangle
    -\langle v_r\rangle \langle B_z\rangle
    \right) \nonumber\\
    &+&
    2\pi r\left(
      \langle v'_z B'_r\rangle
     -\langle v'_r B'_z\rangle
     \right),
\end{eqnarray}
where area $S$ is a circular area with fixed radius $r$ at height $z$, and $\langle\rangle$ shows the azimuthal average.
\cite{2010ApJ...720..233C} suggests that the essential term for collecting the 
magnetic flux at the photosphere is $-\langle v'_rB'_z\rangle$. This is caused by the
mass loss at the highly inclined field (see Fig. 8 in \cite{2010ApJ...720..233C}).\par
\begin{figure}
  \centering
  \includegraphics[width=0.5\textwidth]{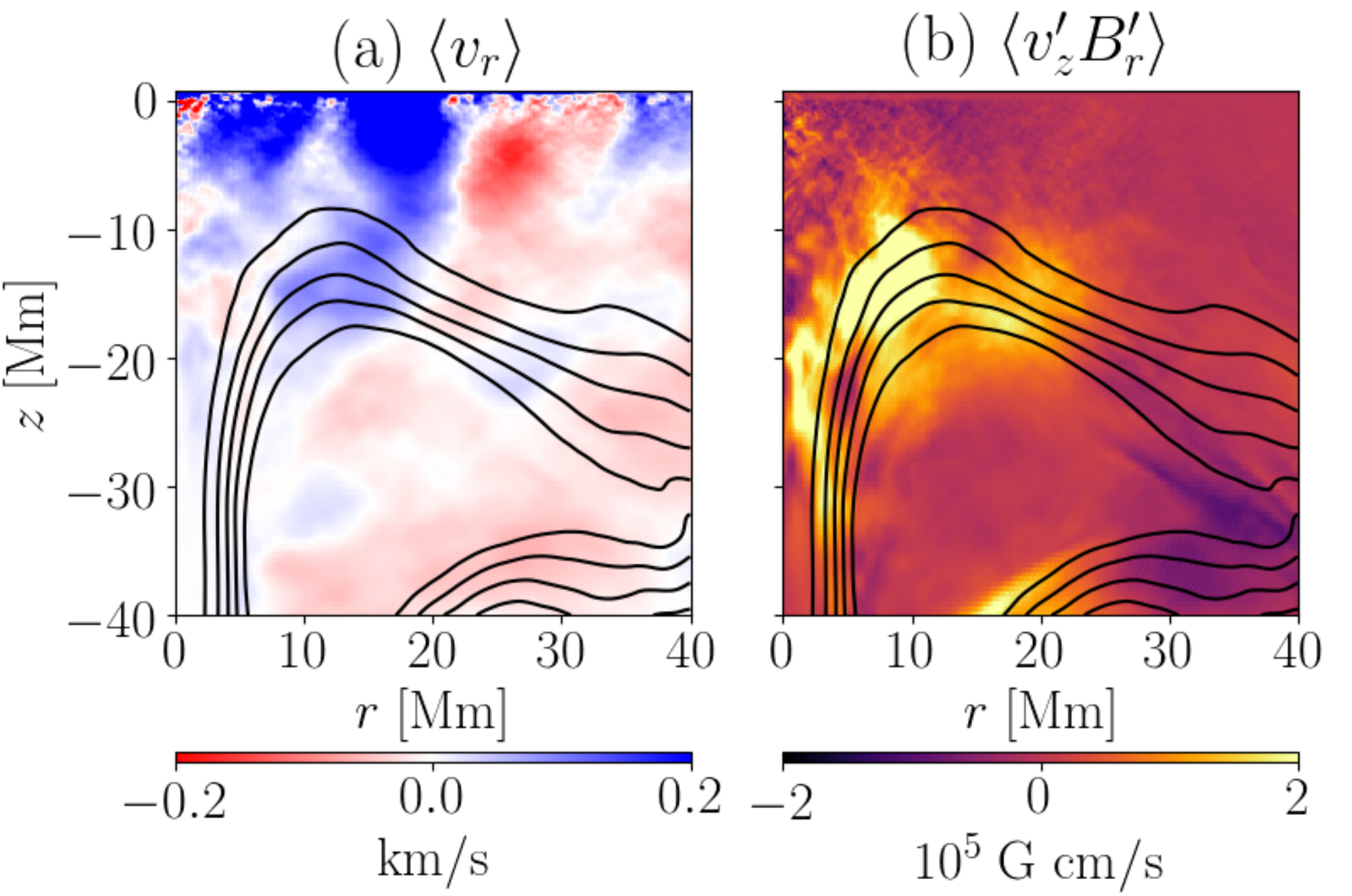}
  \caption{Azimuthally averaged values in the polar coordinate are shown.
  The values are averaged between $t = 33$ and 37 hr.
  (a) The radial velocity 
  $\langle v_r\rangle$
  and (b) one of the induction terms 
  $\langle v'_z B'_r\rangle$ are shown.
  The black contour lines are the same as in Fig. \ref{deep_vertical_flow}.}
  \label{deep_before_formation}
\end{figure}
\begin{figure}
  \centering
  \includegraphics[width=0.3\textwidth]{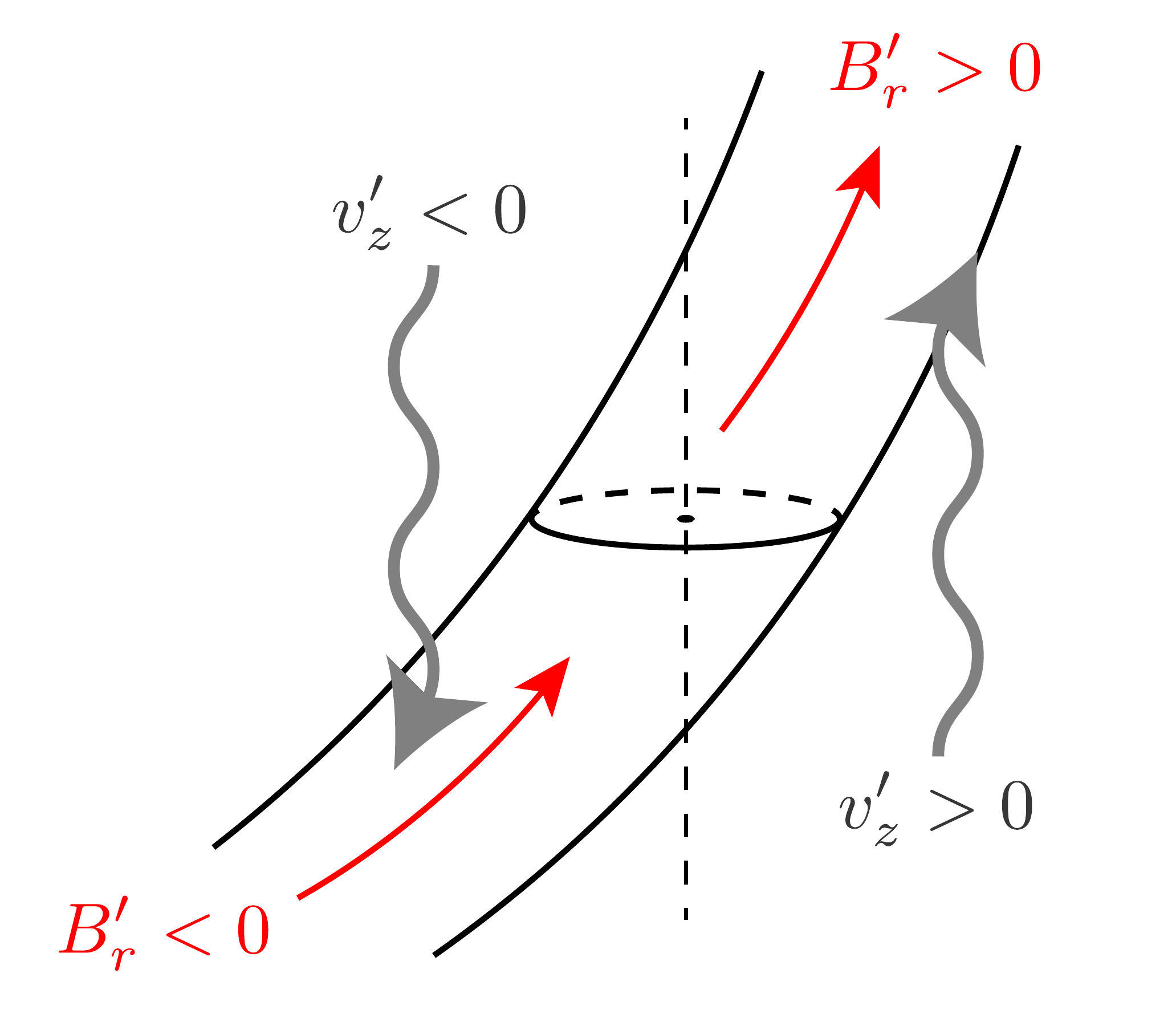}
  \caption{Schematic explaining the positive correlation
  of $\langle v'_z B'_r\rangle$. The red and black arrows show the
  direction of the magnetic field and flow, respectively. The dashed line
  shows the origin of the polar coordinate in this height.}
  \label{exp_positive_correlation}
\end{figure}

Fig. \ref{deep_before_formation} shows the azimuthally averaged radial velocity 
($\langle v_r\rangle$: panel a) and one of the turbulent induction terms
($\langle v'_z B'_r\rangle$: panel b). The definition of the origin of the polar
coordinate is the same as that in Section \ref{flow_tube}.
The values are averaged between $t=33$ and 37 hr, and this range is around the emergence time,
which is before the formation of the sunspot.
At this time, the sunspot location in the
near-surface is filled with the outflow from the center to the outside, which is the
negative effect for the generation of the sunspot. Among the induction terms, the most
important contribution is from $\langle v'_z B'_r\rangle$. In Fig. 
\ref{deep_before_formation}b, this term is positive in the magnetic flux. 
This can be explained with Fig. \ref{exp_positive_correlation}.
The horizontal magnetic field with positive $B_x$ can be regarded as
a negative radial magnetic field ($B'_r<0$) in the downflow side ($v'_z<0$).
In the upflow region ($v'_z>0$), the radial magnetic field is positive ($B'_r>0$).
Both cause a positive correlation, meaning that before the formation of
the sunspot, the 
shear of the downflow and the upflow are the 
most important mechanisms for the generation
of the vertical magnetic flux.\par
\begin{figure}
  \centering
  \includegraphics[width=0.5\textwidth]{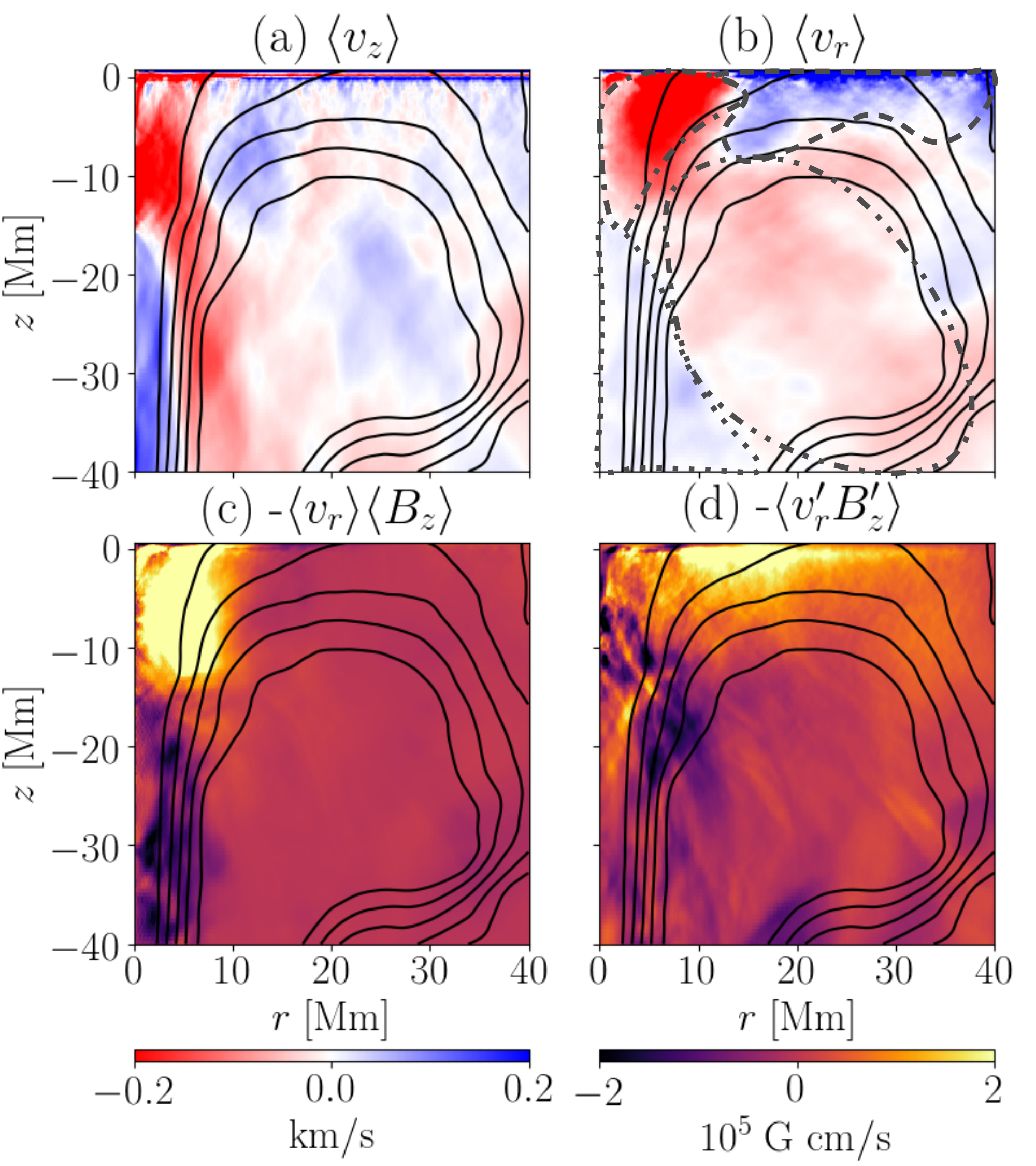}
  \caption{Azimuthally averaged values in the polar coordinate
  during $t = 47$ and 53 hr are shown.
  (a) The vertical velocity $\langle v_z\rangle$,
  (b) the radial velocity $\langle v_r\rangle$,
  (c) the induction terms 
  $-\langle v_r\rangle \langle B_z\rangle $,
  and (d) $-\langle v'_r B'_z\rangle$ are shown.
  The black contour lines are the same as in Fig. \ref{deep_vertical_flow}.
  }
  \label{deep_after_formation}
\end{figure}

Fig. \ref{deep_after_formation} shows the azimuthally averaged flows and
induction terms averaged between $t=47$ and 53 hr, i.e., during the
formation of sunspots. We find a strong coherent
downflow and inflow at the near-surface layer ($z>-15$ Mm)  
in the center 
(Figs. \ref{deep_after_formation}a and b, 
indicated with a dash-dot line). This makes an important
contribution to the collection of the vertical magnetic flux
(Fig. \ref{deep_after_formation}c). This flow is transient and continues less
than 10 hr. While the existence of this downflow and inflow is not reported
in \cite{2010ApJ...720..233C}, \cite{2014ApJ...785...90R} show its importance
(see their Fig. 11). The treatment of the bottom boundary would be important
for the flow for calculations without the deep layer.
\par
In the outer side of the sunspot, we see the Evershed-like outflow 
in the near-surface layer ($-10$ Mm, indicated with a dashed line in Fig. \ref{deep_after_formation}b),
which does not contribute to the loss of the vertical
magnetic flux in this phase, because the Evershed flow
and the magnetic field line
are roughly aligned (Fig. \ref{deep_after_formation}c). In the deeper
layer, the sunspot core is filled with weak outflow, reducing the vertical
magnetic flux in this phase (indicated with a doted line in Fig. \ref{deep_after_formation}b). In the outer area, we also see weak inflow that 
does not play any role in the collection of the vertical flux (indicated with a dash-dot-dot line in Fig. \ref{deep_after_formation}b).\par
The most important contribution of the formation of the sunspot in the outer
region is a turbulent induction term ($-\langle v'_r B'_z\rangle$), which is 
consistent with \cite{2010ApJ...720..233C} and \cite{2014ApJ...785...90R}.
The turbulent induction term ($-\langle v'_r B'_z\rangle$) has
a negative value in the deeper layer ($z<-15$ Mm). Even the sum of all the
induction terms has a negative value in the deep layer during the formation
of the sunspot at the solar surface. Consequently, the decay of the sunspot starts 
during its formation.


\subsection{Sunspot structure}
In this section, we report the overall flow structure of 
the matured sunspot in comparison with the sunspot during 
formation.\par
\begin{figure}
  \centering
  \includegraphics[width=0.5\textwidth]{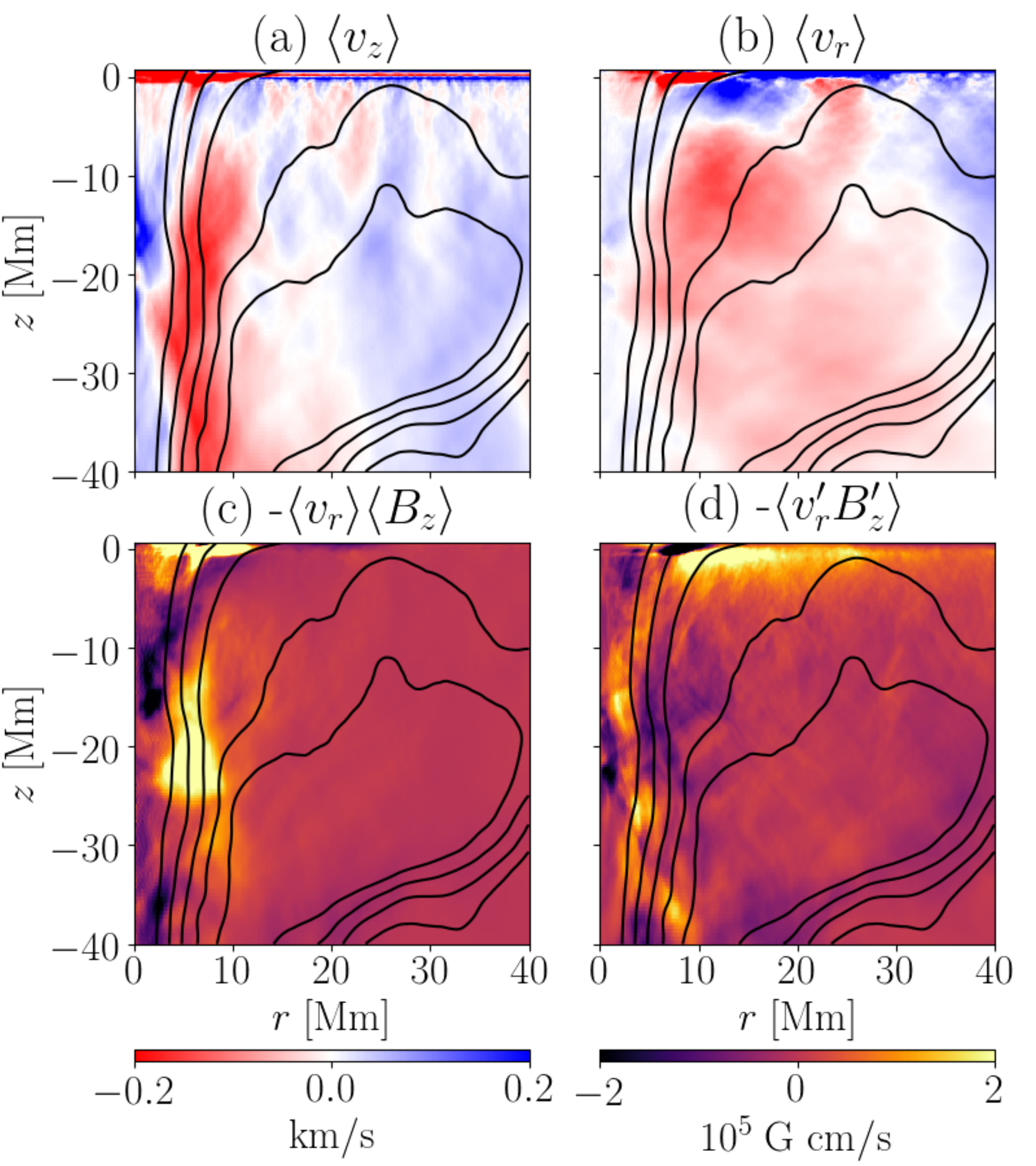}
 \caption{The format of the figure is the same as that in Fig. 
 \ref{deep_after_formation}, but the values are averaged between
 $t=63$ and $69$ hr.}
 \label{deep_mature}
\end{figure}

Fig. \ref{deep_mature} shows the azimuthally averaged flows
and induction terms in the same manner as Fig. 
\ref{deep_after_formation}, but the values are averaged
between $t=63$ to 69 hr, in which the unsigned magnetic
flux has its maximum value. Compared with the sunspot during
formation, the strong coherent downflow and inflow 
in the near-surface layer
disappear (Figs. \ref{deep_mature}a and b), whereas,
weak downflow and inflow are seen in the near-surface layer 
($z>-5\ \mathrm{Mm}$). Nonetheless, the 
weak inflow still contribute to collecting the
vertical magnetic flux in the near-surface layer
 (Fig. \ref{deep_mature}c) . In the deep layer
 ($z=-5$ to $-20$ Mm), the root of the sunspot is filled with
 the outflow, which is connected to the Evershed-like flow
 in the outer area in the near-surface. This outflow in the
 middle layer makes an important contribution to the decay of 
 the sunspot.
 In the deeper layer ($z<-20$ Mm), the center
 of the flux tube is filled with the upflow, as discussed in 
 Section \ref{flow_tube}. This upflow is also connected
 to the outflow in the middle layer and the Evershed-like
 flow in the near-surface. We see coherent inflow in the
 outside the flux tube (Fig. \ref{deep_mature}b), which
 is connected to the downflow in the outer layer of the flux
 tube in the deeper layer. This coherent inflow makes a
 significant contribution to the collection of the magnetic
 flux, i.e., the maintenance of the sunspot.
 The turbulent induction term 
 $-\langle v'_r B'_z\rangle$ still makes a significant
  contribution to the maintenance of the sunspot in the
  near-surface layer, even if the sunspot is matured.\par
\begin{figure}
  \centering
  \includegraphics[width=0.5\textwidth]{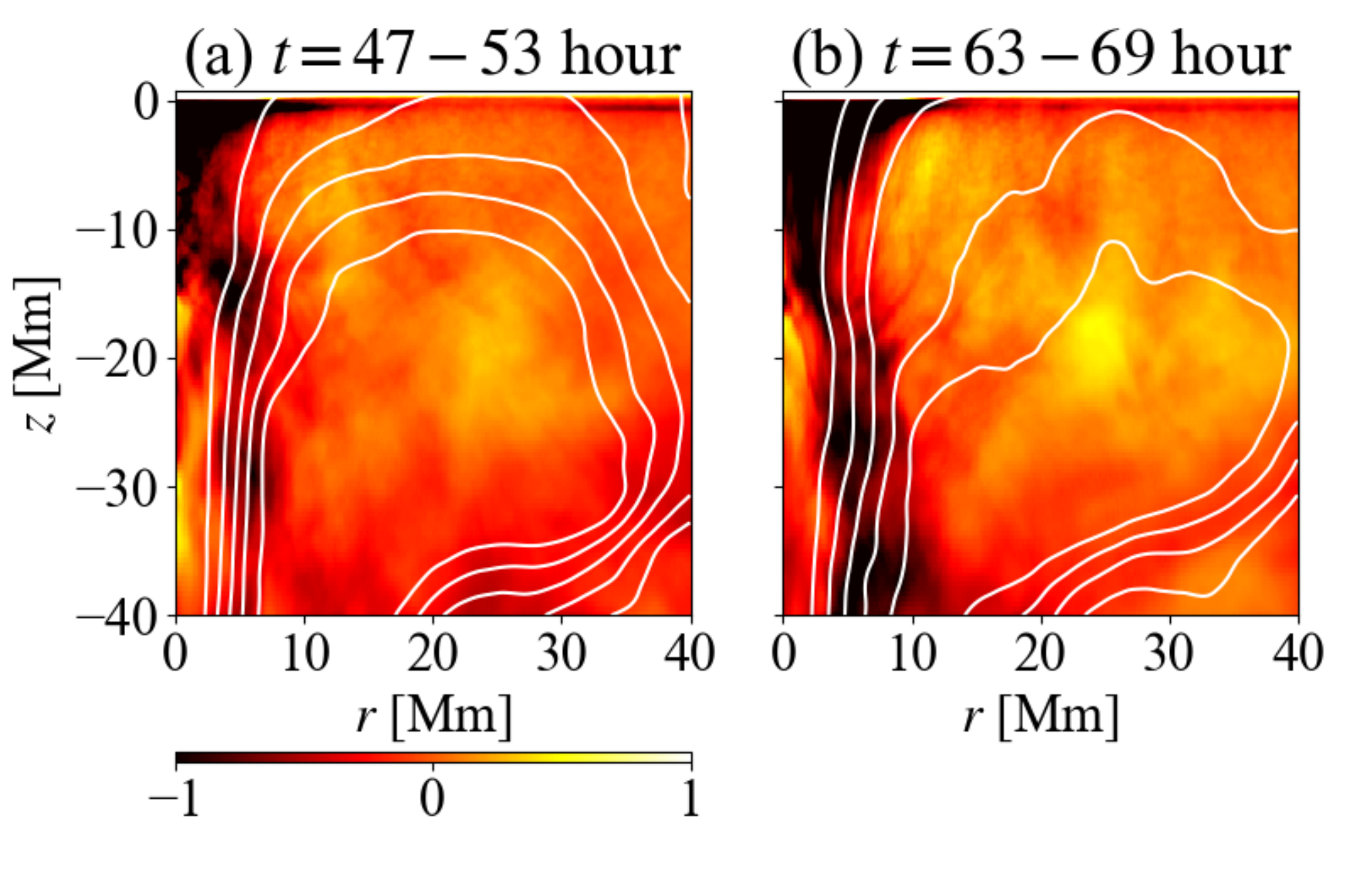}
  \caption{Azimuthally averaged normalized temperature
  $\tilde{T}$ is shown. Panels a and b show the average 
  value in $t=47$ and 53 hr (during the sunspot formation)
  and $t=63$ and 69 hr (matured sunspot), respectively.}
 \label{deep_temperature}
\end{figure}

  Fig. \ref{deep_temperature} shows the azimuthally averaged
  normalized temperature $\tilde{T}$ averaged between
  $t=47$ and 53 hr (panel a, during the formation)
  and $t=63$ and 69 hr (panel b, matured sunspot), where
  $\tilde{T}=(T-\langle T\rangle_\mathrm{hd})/T_\mathrm{rms(hd)}$.
  As the formation process of the
  sunspot proceeds, the low-temperature region expands
  because of the suppression of the convective energy transport
  and the radiative cooling at the photosphere.
  In both cases, we see a high-temperature region in the 
  center of the flux tube below a depth of $-15$ Mm. 
  While the phase of the sunspot formation is different
  in panels a and b, the depth where the temperature
  changes from high to low does not change. In the
  matured sunspot, the coherent low-temperature region continues to
  a depth of 40 Mm.
  
\section{Summary and Discussion}\label{summary}
In this study, we perform a numerical experiment of 
the flux emergence in an unprecedentedly deep domain that
covers the whole convection zone from the base to the surface.
The main purpose of this setting is to minimize the influence of 
the bottom boundary condition on the rising of the flux 
tube, formation of the sunspot, and internal structure
of the generated sunspot.\par
The main findings of this study are summarized as follows:
\begin{itemize}
  \item The rising speed of the flux tube tends to be larger than 
  the typical upward convection velocity because of the low density
  caused by the magnetic pressure and suppression of 
  mixing.
  \item The rising speed of the flux tube exceeds 
  $250\ \mathrm{m\ s^{-1}}$ at a depth of 18 Mm, while no clear
  evidence of the divergent flow 3 hr before the
  emergence at the solar surface is seen.
  \item Initially, the root of the flux tube is filled with
  the downflows, and then the upflow fills the center of the flux tube
  during the formation of the sunspot because of the low density.
  \item The essential mechanisms for the formation of the
  sunspot are the coherent inflow in the central region and
  the turbulent correlation $-\langle v'_r B'_z\rangle$ in 
  the outer region.
  \item The low-temperature region is extended to
  a depth of at least of 40 Mm in the matured sunspot, with the high-temperature
  region in the center of the flux tube.
\end{itemize}

\begin{figure}
  \centering
  \includegraphics[width=0.5\textwidth]{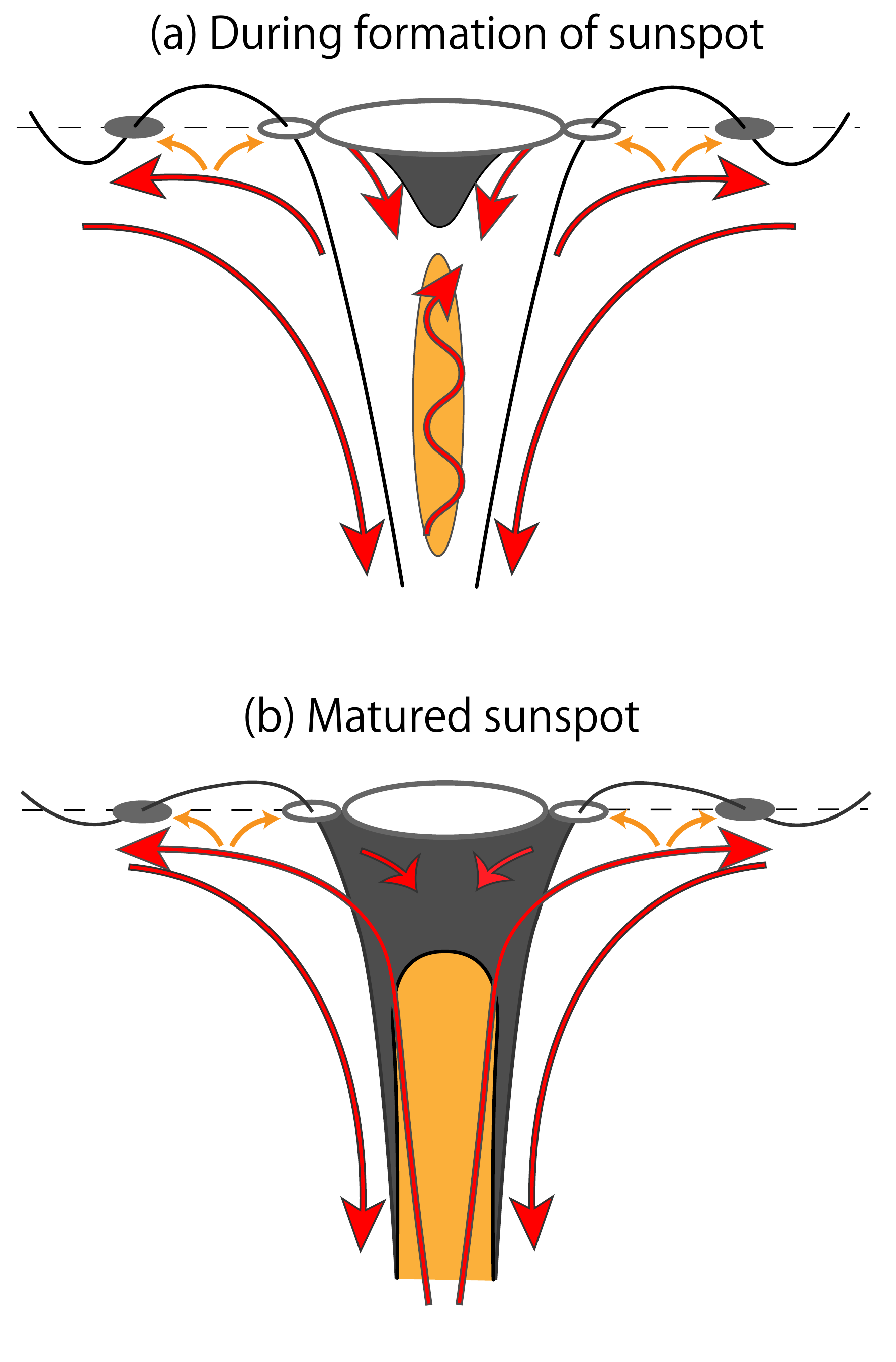}
  \caption{The summary of the sunspot structure during
  formation (panel a) and the matured sunspot (panel b).
  The red arrows show the coherent mean flow, and the orange arrows
  show the turbulent flow, which contributes to the collection of
  the magnetic flux. The black and orange areas correspond to
  low and high temperatures, respectively.
  }
  \label{exp_sunspot}
\end{figure}

Fig. \ref{exp_sunspot} summarizes the flow and temperature structure
of the sunspot during the formation (panel a) and matured phase
(panel b). A large difference in the sunspot structure between the two phases
is the strong coherent downflow/inflow to the center of the sunspot
down to a depth of 15 Mm during its formation. This downflow
is significantly weakened in the matured sunspot and extends only 
5 Mm further. The other difference between the two phases is the temperature
structure. During the formation, the low-temperature region is
mainly seen up to a depth of 10 Mm (some cooling is still seen at a depth of $30$ Mm);
a coherent low-temperature region is seen down to a depth of 40 Mm surrounding the
high-temperature region in the center of the flux tube.

\subsection{Reduced Speed of Sound Technique}
\begin{figure}
  \centering
  \includegraphics[width=0.5\textwidth]{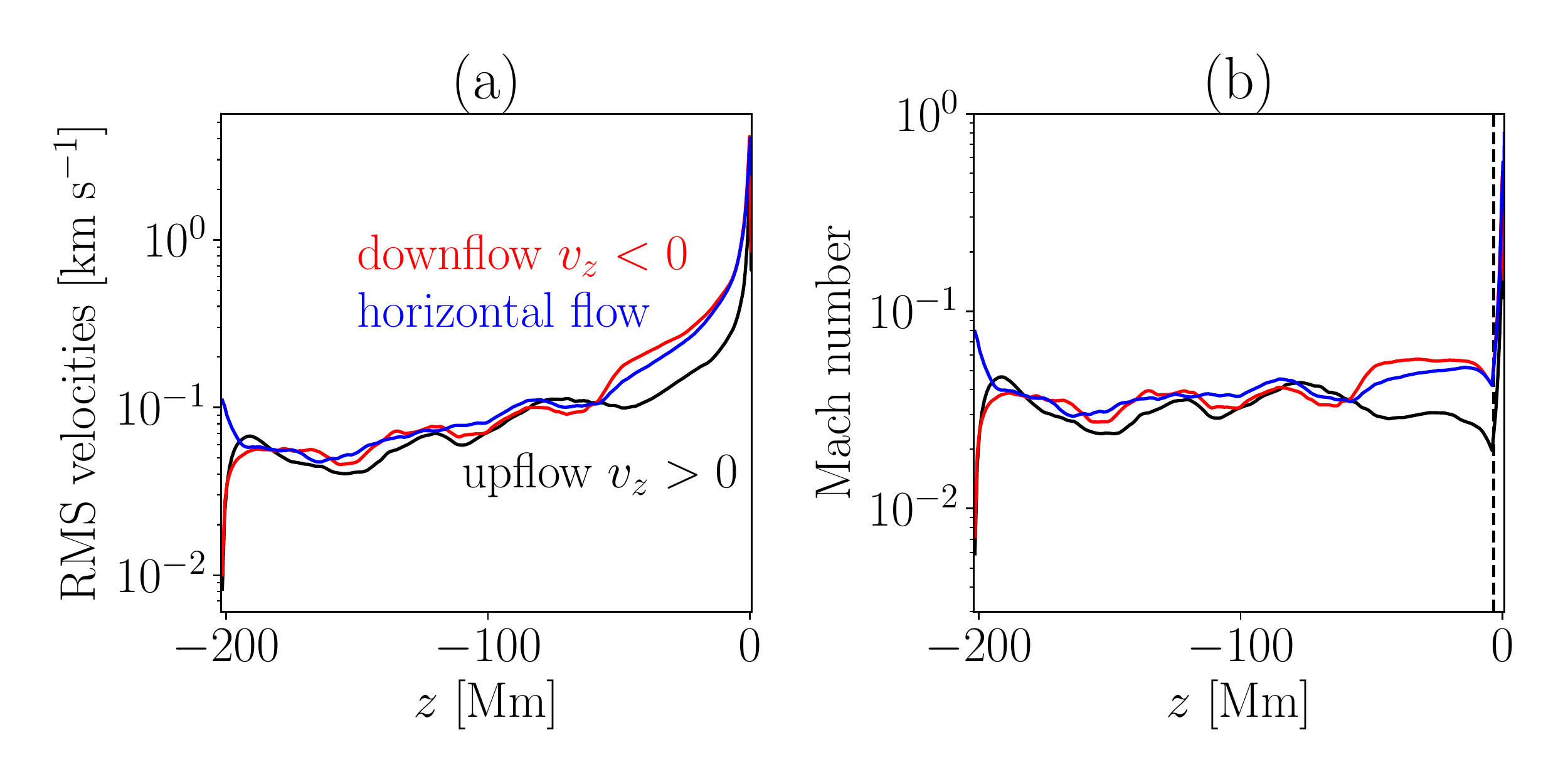}
  \caption{
  (a) The RMS velocity in the hydrodynamic calculation without the magnetic field and (b) the Mach number estimated with the reduced speed of sound are shown.
  The black, red, and blue lines show the results with upflow, the downflow and the horizontal flow, respectively. Below the location indicated with the dashed line in the panel b, the RSST is used ($\xi>1$).
  }
  \label{rms_rsst}
\end{figure}

As discussed in \cite{2012A&A...539A..30H}, we need to keep the Mach number smaller than 0.7 in order to avoid the influence of the RSST. In this study, we keep the Mach number smaller than 0.1 when the RSST is used. Fig. \ref{rms_rsst} shows the RMS velocity in the hydrodynamic calculation without the magnetic field (panel a) and the Mach number estimated with the reduced speed of sound (panel b). The Mach number is always smaller than 0.1 at locations where the RSST is used (indicated with the dashed line in Fig \ref{rms_rsst}b). Phenomena related with the magnetic field, i.e., the flux emergence and sunspot formation, occurs in a similar time scale as the thermal convection. For example, Fig. \ref{tube_internal_flow} shows that the up- and downflow are typically 200--300 $\mathrm{m\ s^{-1}}$, where the reduced speed of sound at z=-30 Mm is about 4 $\mathrm{km\ s^{-1}}$. Thus the Mach number is smaller than 0.1 and we do not expect any significant influence of the RSST on the result shown in this paper.
\subsection{Computational domain}
In this study, we prepare an unprecedentedly deep computational 
domain to minimize the bottom boundary influence. This 
also has a possible drawback on the convection structure. It is 
known that numerical simulations tend to overestimate the
large-scale (> 30 Mm) convection energy in the deep convection 
zone \citep{2012PNAS..10911928H,2014ApJ...793...24L}. To
reduce the power on a larger scale than the supergranulation,
we intentionally prepare the small box in the horizontal direction 
of 100 Mm \citep[see also][]{2019ApJ...886L..21T}. This setting
causes another side effect. Because of the insufficient horizontal
box size and the periodic boundary condition, a generated 
sunspot is close to the sunspot on the other side.
We expect that the calculation results in this study are
not an evolution of the single isolated sunspot pair, but rather, the sequence of 
the sunspot pairs. This fact mainly influences the decay of the 
sunspot, and we do not discuss this effect in detail in this study.

\subsection{Emergence rate}
In this study, the emergence rate of the magnetic flux
is $9\times10^{20}\ \mathrm{Mx\ hr^{-1}}$, which is 
slightly larger than the 95\% confidence interval 
($7\times10^{20}\ \mathrm{Mx\ hr^{-1}}$, 
see Section \ref{overall_evolution}) suggested by
\cite{2019ApJ...871..187N}. There are several
possible reasons for this discrepancy.
The emergence rate would be determined by the rising speed and 
the structure of the magnetic flux tube.
A fast-rising and intense structure of the flux tube tends
to cause a high emergence rate. In this study, we find that
the rising speed of the flux tube is not purely determined by
the thermal convection; the magnetic effect is also important.
Thus, there would be an appropriate initial setting of the magnetic flux tube 
for reproducing the emergence rate consistent with the observation. 
Parameter surveys for the initial setting of the
flux tube should therefore be carried out in a future study.\par
By contrast, the number of observations is insufficient to 
conclude that our emergence rate is larger than reality. 
There are only several estimations of the emergence rate for the magnetic 
flux larger
than $10^{22}\ \mathrm{Mx}$ \citep{2014ApJ...794...19T}. For example,
\cite{2017RNAAS...1...24S} show $1.12\times10^{21}\ \mathrm{Mx\ hr^{-1}}$ for
an instantaneous emergence rate. More observations are needed to identify
the most probable flux tube structure in the deep convection zone.

\subsection{Comparison with previous studies}
\subsubsection{Divergent flow prior to flux emergence}
In this study, we find that the rising speed of the flux tube exceeds 
$250\ \mathrm{m\ s^{-1}}$ at a depth of 18 Mm, while we do not see any clear evidence of 
the divergent flow 3 hr before the emergence time.
This is inconsistent with \cite{2016SciA....2E0557B}, who 
argue that the rising speed should not be larger than 
$150\ \mathrm{m\ s^{-1}}$ to have divergent flow at that time.
This discrepancy can be  caused by the settings in previous studies
\citep{2010ApJ...720..233C,2014ApJ...785...90R}. In those studies,
a magnetic flux tube has to be inserted kinematically from the
bottom boundary because of the shallow calculation box.
The speed of the rising flux tube is determined at the bottom 
boundary, and most of the flux tube has upward velocity.
In this study, the flow structure of the flux tube is physically 
determined. The rising part of the flux tube has upflow, and 
its root initially has downflow. This natural determination
of the rising speed would reduce the influence on the 
photosphere, i.e., the divergent flow. The constraints on the 
rising flux tube should be investigated in a deeper computational
box, where the interaction of the convection and the magnetic field
automatically determines the dynamics of the flux tube.

\subsubsection{Upflow in flux tube}

The other difference from previous studies is the upflow in the
central region of the flux tube.
\cite{2017ApJ...846..149C} found that the flux tube of the sunspot
is filled with coherent downflow (see Fig. 11 in their work).
In their study, they use the data from the dynamo calculation
\citep{2014ApJ...789...35F}. They also find that the downflow
region is the place for the sunspot because the converging flow to
the downflow collects the magnetic flux. The flow structure does
not change even after the sunspot is formed because the boundary
condition is provided. In our study, we do not expect any 
boundary influence on the evolution of the flow in the flux tube,
and the upflow in the central region with low density is a natural
consequence of the magnetohydrodynamic process.
\par
\cite{2003ApJ...591..446Z} show the internal flow structure 
beneath the sunspot using the local helioseismology.
They find converging motion with the downflow in the near-surface
layer (depths of $0-3$ Mm) and diverging motion with the upflow 
in the deeper layer (depths of $9-12$ Mm). Our result for matured sunspots has
a consistent flow structure (see Fig. \ref{deep_mature}).
In addition, Fig. 4 of \cite{2003ApJ...591..446Z}
shows that the downflow and upflow are surrounded by
the upflow and downflow, respectively, but this information is not 
explicitly mentioned in the paper. This feature is also
consistent with our calculation.

\subsection{Future perspective}
Previously, flux emergence simulations in the deep convection zone with
a thin-flux tube \citep{1993A&A...272..621D,1995ApJ...452..894M}, 
anelastic approximation \citep{2008ApJ...676..680F,2011ApJ...741...11W},
RSST \citep{2012A&A...548A..74H}, and 
simulations around the solar surface
\citep{2010ApJ...720..233C,2014ApJ...785...90R,2017ApJ...846..149C} were
almost completely separated. Our approach using the R2D2 code sheds light
on connecting these two regions. This is an integral part of both
the solar dynamo and the formation of the sunspot.
Even in the simulation in this study with the computational domain of the
whole convection zone, we still assume the initial magnetic
flux tube condition because we have a very low vertical resolution
in the deep convection zone ($\sim900\ \mathrm{km}$).
To maintain the magnetic flux of the magnetic flux tube during
the rising process from the base of the convection zone to the surface,
we need to prepare a high resolution even in the deep convection zone
\citep{2006A&A...451..303C}. By using the current numerical resources,
we are unable to achieve this purpose; however, we will soon be able to
perform these types of calculations using next-generation
supercomputers, such as Fugaku in Japan. In future studies, we plan to connect dynamo
calculation and sunspot formation simulation in a calculation.
This approach is expected to reveal the formation process of sunspots in 
a more self-consistent manner.

\section*{Acknowledgments}
The authors would like to thank the anonymous referee for the great suggestions.
The authors are grateful to Kousuke Namekata for providing the detailed
relation of the sunspot flux and the emergence rate.
The results were obtained using the K computer at the RIKEN  Advanced Institute for Computational Science 
(proposal Nos. hp190070, hp190183, hp180042, and ra000008). 
This work was supported by MEXT/JSPS KAKENHI grant Nos. JP19K14756, JP18H04436, JP16K17655, and JP15H05816.
This research was supported by MEXT as Exploratory Challenge on Post-K Computer
(Elucidation of the Birth of Exoplanets [Second Earth] and the Environmental 
Variations of Planets in the Solar System) and the Project for Solar-Terrestrial Environment Prediction. 

\appendix

\section{Background stratification}\label{background}
For the background stratification of $\rho_0$, $p_0$, $T_0$, and
the other related variables, 
we use Model S \citep{1996Sci...272.1286C}.
Model S does not solve the entropy equation, and the entropy
profile is not smooth.
By contrast, in this study, the entropy equation is solved,
and the tiny variation of 
the entropy is essential for the thermal convection property.
To overcome this difficulty, we recalculate
the hydrostatic equation with the help of the Model S.
The hydrostatic equation is as follows:
\begin{eqnarray}
 \frac{d p_0}{dz} = -\rho_0 g.
\end{eqnarray}
We adopt the value of the gravitational acceleration
$g$ from the Model S, and we need to specify the
distribution of the temperature or the entropy to
relate the pressure $p_0$ and the density $\rho_0$.
We divide the solar convection zone into two parts at
$z=-3.5\ \mathrm{Mm}$. In the deep part, the entropy gradient is small,
and we can regard it as adiabatic stratification.
Thus, in this region, we adopt a constant value of the entropy
at $z=-3.5\ \mathrm{Mm}$ in Model S.
\par
In the upper part of the convection zone ($z>-3.5\ \mathrm{Mm}$),
the entropy gradient is significantly large. 
To obtain the smooth profile of the entropy gradient
around the photosphere, we follow the following procedure:
\begin{enumerate}
 \item Calculate the entropy gradient $ds_0/dz$
 using the entropy of the Model S.
 \item Use the Savitzky--Golay filter for the entropy gradient
 to obtain a smooth profile.
 \item Solve the hydrostatic equation using the filtered profile.
\end{enumerate}
\begin{figure}
  \centering
  \includegraphics[width=0.5\textwidth]{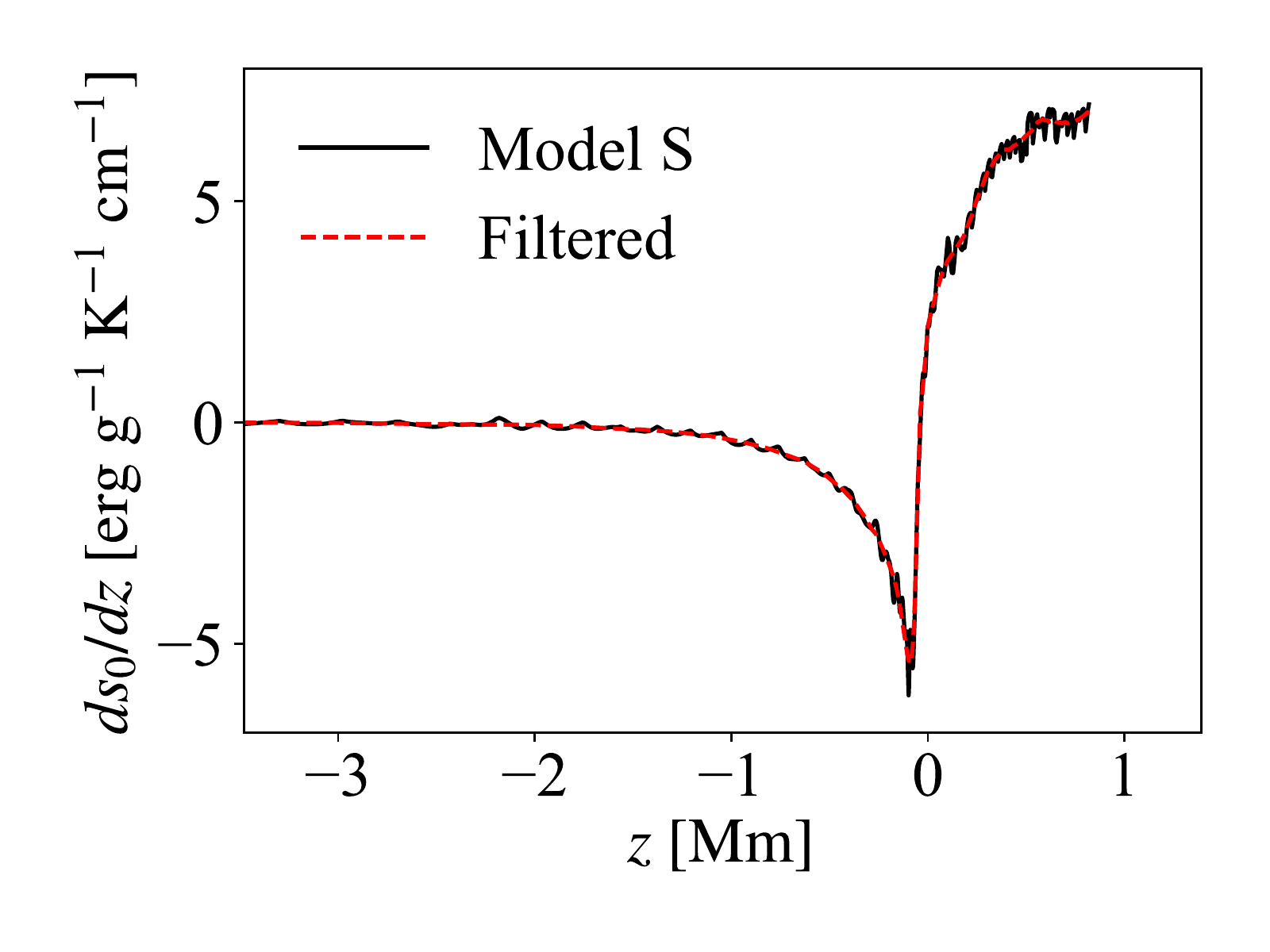}
  \caption{Entropy gradient in the upper part of the convection zone.
 The black line shows the original value in Model S, and the red
 line shows the filtered value.
   \label{filter_entropy}}
\end{figure}
The black line in Fig. \ref{filter_entropy} shows the calculated
entropy gradient in Model S. To 
reduce the jaggy feature in the entropy gradient,
we use the Savitzky--Golay filter.
The red line in Fig. \ref{filter_entropy} shows the filtered
entropy gradient
\par
In addition, the top boundary of
the Model S is about 500 km above the photosphere.
We extend the stratification
with the gravitational acceleration 
$g(z)\propto \left(z+R_\odot\right)^{-2}$
and constant temperature, where we again note that $z=0$
corresponds to the surface of the sun.

\section{Fourth-order derivative for inhomogeneous grid}
\label{ihomogeneous_grid}
In this study, we adopt inhomogeneous grid spacing in 
the vertical direction
$z$. To maintain the accuracy of the spatial
derivative of the quantities, we adopt a general fourth-order
formulation for inhomogeneous grid spacing as follows:
\begin{eqnarray}
 z_{i+2} - z_i &=& a, \\
 z_{i+1} - z_i &=& b, \\
 z_{i} - z_{i-1} &=& c, \\
 z_{i} - z_{i-2} &=& d.
\end{eqnarray}

The first derivative of a quantity ($q$) can be expressed as follows:
\begin{eqnarray}
 \left(\frac{\partial q}{\partial z}\right)_i &=&  
 \frac{c d-b (c+d)}{(a-b) (a+c) (a+d)} q_{i+2}
 + \frac{a (c+d)-c d}{(a-b) (b+c) (b+d)} q_{i+1}  \nonumber\\
 &+&  \frac{a (d-b)+b d}{(a+c) (b+c) (c-d)} q_{i-1}
  + \frac{a (b-c)-b c}{(c-d) (a+d) (b+d)} q_{i-2}.\nonumber\\
\end{eqnarray}

\section{Radiation transfer}\label{radiation}
\begin{figure}
  \centering
  \includegraphics[width=0.5\textwidth]{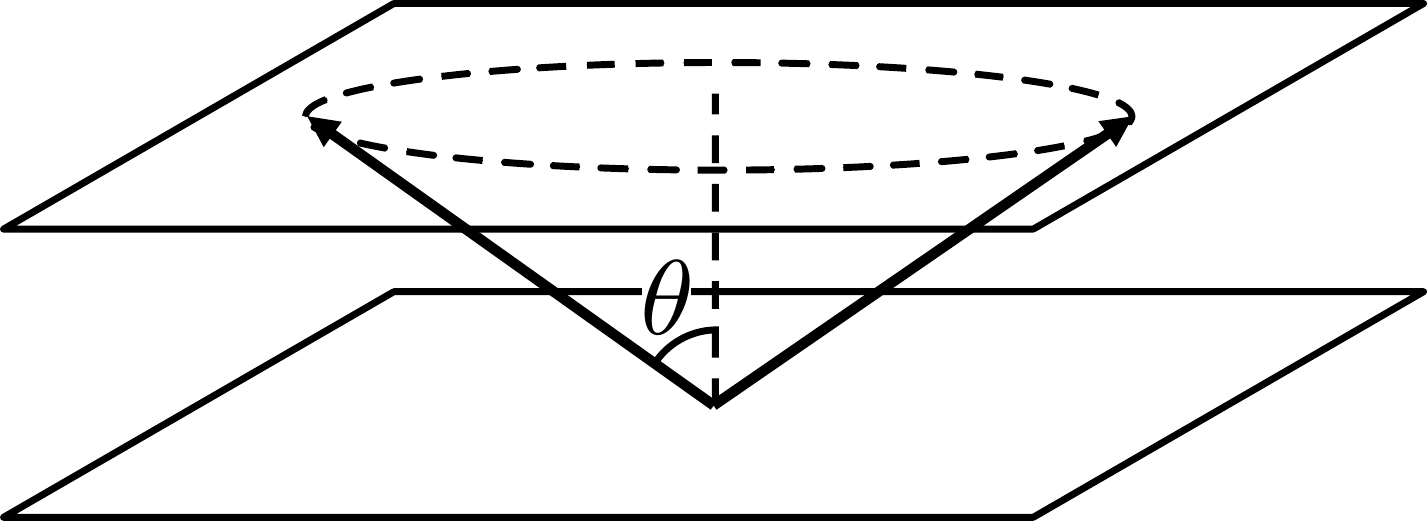}
  \caption{Schematic of our treatment of the radiative
  transfer. The arrow shows a ray inclined from the vertical
  axis with the angle of $\theta$. We assume the azimuthal homogeneity
 of the radiative intensity and that the radiative energy is transported
 only upward and downward.
   \label{rad}}
\end{figure}

To evaluate the heat and the cooling of the radiation
($Q$ in Eq. (\ref{entropy})), we solve the radiation transfer equation.
\begin{eqnarray}
\frac{\partial I}{\partial \tau} = - I + S,
\end{eqnarray}
where $I$ is the radiative intensity, $S=\sigma T^4/\pi$
is the source function, and $\sigma$ is the Stefan--Bolzman constant.
In this study, we use the Rosseland mean opacity for
the gray radiation transfer.
Every quantity in the MHD  equations is defined in the cell center, and
the value at the cell surface is needed for the radiation transfer.
To this end,
we adopt the linear interpolation for the logarithmic value, i.e.,
\begin{eqnarray}
  q_{i+1/2} = \exp\left[\frac{\ln\left(q_i\right) + 
  \ln\left(q_{i+1}\right)}{2}\right].
\end{eqnarray}
In this study, we treat only vertically upward and downward radiation
energy transport.
We solve rays inclined to the vertical axis (Fig. \ref{rad}).
The inclination is expressed as $\mu=\cos\theta$ with $\mu=1/\sqrt{3}$. 
Then, the intensity is azimuthally averaged for evaluating
the radiative heating ($Q$).
For example, we explain the upward radiation transfer from
$z_{i-1/2}$ to $z_{i+1/2}$, where 
$z_{i+1/2}=\left(z_{i+1}+z_{i}\right)/2$.
As a first step, we evaluate the optical depth between $z_{i-1/2}$ and
$z_{i+1/2}$ along a ray, which is expressed as follows:
\begin{eqnarray}
  \Delta \tau = \frac{1}{\mu}\int_{z_{i-1/2}}^{z_{i+1/2}} 
  \rho \kappa dz.\label{define_tau}
\end{eqnarray}
We assume that the logarithmic value of $\alpha=\rho\kappa$, where $\kappa$ is the opacity, is a linear
function in the space $z$, i.e.,
\begin{eqnarray}
  \ln \alpha (z) = \ln \alpha_{i-1/2} + \frac{\ln \alpha_{i+1/2}-
  \ln \alpha_{i-1/2}}{z_{i+1/2}-z_{i-1/2}}\left(z-z_{i-1/2}\right).
\end{eqnarray}
Then, Eq. (\ref{define_tau}) can be integrated analytically, and
the solution is
\begin{eqnarray}
  \Delta \tau =
  \frac{\left(\alpha_{i+1/2}-\alpha_{i-1/2}\right)
        \left(z_{i+1/2}-z_{i-1/2}\right)}{\ln \alpha_{i+1/2}
        -\ln \alpha_{i-1/2}}. \label{define_tau2} \nonumber \\    
\end{eqnarray}
Eq. (\ref{define_tau2}) includes a singularity at
$\alpha_{i-1/2} = \alpha_{i+1/2}$. To avoid the singularity,
 we use
a different expression for $\Delta \tau$ using the Taylor expansion 
around
$\alpha_{i-1/2} = \alpha_{i+1/2}$ as
\begin{eqnarray}
  \Delta \tau =
  \left[
  \frac{\alpha^2_{i-1/2}}{3\alpha_{i-1/2}-\alpha_{i+1/2}} +
  \frac{\alpha^2_{i+1/2}}{3\alpha_{i+1/2}-\alpha_{i-1/2}}
  \right]
  \left(
    z_{i+1/2} - z_{i-1/2}
  \right).\nonumber\\
\end{eqnarray}
When the upward radiation transfer is considered,
the intensity at the cell surface ($I_{i+1/2}$) is determined by the
intensity in the downstream cell surface ($I_{i-1/2}$)
with the formal solution of the radiation transfer equation.
\begin{eqnarray}
  I_{i+1/2} &=& I_{i-1/2}\exp\left(-\Delta \tau\right) 
  \nonumber\\
  &&+ \int_0^{\Delta \tau} 
  S\left(\Delta\tau'\right)\exp\left(-\Delta \tau +
  \Delta \tau'\right)d\left(\Delta \tau'\right), \label{ref}
\end{eqnarray}
To solve Eq. (\ref{ref}) analytically, we follow a similar 
procedure for the evaluation of the optical depth.
While the linear function \citep{2005A&A...429..335V} and
the second-order B{\'e}zier curve \citep{2003ASPC..288....3A} 
are used for
the source function in previous studies, we adopt the linear 
function
for the logarithmic values, i.e.,
\begin{eqnarray}
    \ln S\left(\Delta \tau'\right) = \ln S_{i-1/2}
    +\frac{\ln S_{i+1/2} - \ln S_{i-1/2}}{\Delta \tau}\Delta\tau'.
\end{eqnarray}
Then, Eq. (\ref{ref}) is solved analytically as
\begin{eqnarray}
    I_{i+1/2} = I_{i-1/2}\exp\left(-\Delta \tau\right)
    + \Delta \tau
    \frac{S_{i+1/2} - S_{i-1/2}\exp\left(-\Delta\tau\right)}
    {\ln S_{i+1/2}-\ln S_{i-1/2} + \Delta \tau}.
\end{eqnarray}
The downward radiation can be calculated using the same method 
in the opposite
direction.\par
When the upward and downward intensities on the cell surface
($I_{\mathrm{(up)}i+1/2}$, and $I_{\mathrm{(dw)}i+1/2}$, respectively)
are evaluated, the radiative heating is calculated 
in two ways depending on
the optical depth. For the small optical depth $\tau$, 
we use the expression
\begin{eqnarray}
  Q_{(\mathrm{J})i} &=& 4\pi \rho_{i}\kappa_{i}
  \left(J_i - S_i\right), \\
  J_i &=& \frac{I_{\mathrm{(up)}i+1/2}+I_{\mathrm{(up)}i-1/2}+
                I_{\mathrm{(dw)}i+1/2}+I_{\mathrm{(dw)}i-1/2}}{4}.\nonumber\\
\end{eqnarray}
For the large optical depth,
\begin{eqnarray}
  Q_{(\mathrm{F})i} &=&
    -\frac{F_{\mathrm{(rad)}i+1/2}-F_{\mathrm{(rad)}i-1/2}}{\Delta z},\\
    F_{\mathrm{(rad)i+1/2}} &=&
    2\pi \mu \left( I_{\mathrm{(up)}i+1/2} - I_{\mathrm{(dw)}i+1/2}\right).
\end{eqnarray}
We switch these expressions around $\tau = 0.1$, as follows:
\begin{eqnarray}
  Q_i = Q_{\mathrm{(J)}i}\exp\left(-\frac{\tau}{\tau_0}\right)
  +Q_{\mathrm{(F)}i}\left[1-\exp\left(-\frac{\tau}{\tau_0}\right)\right],
\end{eqnarray}
where $\tau_0=0.1$ and the validation of this method is detailed in our previous
publication \citep{2019SciA....eaau2307}.







\bsp	
\label{lastpage}
\end{document}